\documentstyle[aps,prl,eqsecnum]{revtex}
\begin{document}
\draft
\twocolumn[\hsize\textwidth\columnwidth\hsize\csname
@twocolumnfalse\endcsname
\date{May 9, 2000}
\title{Anholonomic Soliton--Dilaton and Black Hole
       Solutions in General Relativity}
\author{Sergiu I. Vacaru}
\address{Institute of Applied Physics, Academy of Sciences,\\
5 Academy str., Chi\c sin\v au\ MD2028, Republic of Moldova \\ {--- } \\
 Electronic address: vacaru@lises.asm.md \\ {--- } }
\maketitle
\vskip0.3cm
\begin{abstract}
 A new method of construction of integral varieties of Einstein equations
 in three dimensional (3D) and 4D gravity is presented whereby, under
 corresponding   redefinition of physical values with respect
 to  anholonomic frames  of reference  with associated nonlinear connections,
  the structure of gravity  field equations is   substantially simplified.
  It is shown that there are 4D solutions of Einstein equations  which are
  constructed as  nonlinear superpositions of  soliton solutions of 2D
  (pseudo) Euclidean  sine--Gordon equations (or of Lorentzian black holes
  in  Jackiw--Teitelboim dilaton gravity). The Belinski--Zakharov--Meison
 solitons for vacuum gravitational field equations are generalized
 to various cases of two and three coordinate dependencies, local anisotropy
 and matter sources. The general framework of this study is based on
 investigation of anholonomic soliton--dilaton black hole  structures in
 general relativity. We prove that  there are possible  static and dynamical
  black hole, black torus and disk/cylinder like solutions (of non--vacuum
 gravitational field equations) with horizons being parametrized
 by hypersurface equations of rotation ellipsoid, torus, cylinder and another
 type configurations. Solutions describing locally anisotropic variants of
 the  Schwarzschild--Kerr (black hole), Weyl (cylindrical symmetry) and
 Neugebauer--Meinel (disk) solutions with anisotropic variable masses,
 distributions of matter and  interaction constants are  shown
  to be contained in Einstein's gravity.
 It is demonstrated in which manner locally anisotropic
 multi--soliton--dilaton--black  hole  type  solutions can be generated.
 \end{abstract}

\pacs{98.80.Cq, 04.50.+h, 98.80.Hw, 02.90.+p, 04.70.Bw, 11.10Kk
 \qquad E--preprint:\ {\bf gr--qc/0005025} }
\vspace{0.5cm}  ] %\tableofcontents

\section{Introduction}

\subsection{Soliton--black hole solutions in modern gravity and string/brane
theories}

One of the main ingredients of the last developments in string theory and
gravity is the investigation of connections between black holes and
non--perturbative structures of string theory such as
Bogomol'nyi--Pasad--Sommerfeld (BPS) solitons or D--branes \cite
{duff,strominger}. New ways to address and propose solutions of old and new
fundamental problems of black hole physics are opened via explanation of
black hole thermodynamics in terms of microscopic string and membrane
physics.

Similar fundamental problems of black hole physics have been recently
analyzed in the recent literature using low--dimensional gravity models
(see, for instance, Ref. \cite{strominger1}). Dilaton gravity theories in
two spacetime (2D) dimensions have been used as theoretical laboratories of
studding 4D black hole physics.

One of 2D theories of particular interest is the so--called
Jackiw--Teitelboim theory (JT) \cite{jackiw} because of its connection to
the Liouville--Polyakov action in string theory. The black hole solutions to
JT--gravity are dimensional reductions of the Banados--Teitelboim--Zanelli
(BTZ) black hole solutions \cite{btz} and exhibit usual thermodynamic
properties with black hole entropy \cite{lemos}. Such JT black hole
solutions describe spacetimes of constant curvature.

The aim of this paper is to develop and apply a new method of construction
of 4D solutions of the Einstein equations by considering nonlinear
superpositions of 2D and 3D soliton--dilaton--black hole metrics which are
treated as some non--perturbative structures in general relativity.

\subsection{Locally anisotropic soliton, black hole and disk/cylinder like
solutions}

Mathematicians (see, for instance, the review \cite{solitons}) have
considered for a long time the relationship between Euclidean, $\epsilon =1$
(pseudo--Euclidean, or Lorentzian, $\epsilon =-1),$ 2D metrics
\begin{equation}
\label{soliton1}ds^2=\epsilon \sin ^2\left( \frac \phi 2\right) dt^2+\cos
^2\left( \frac \phi 2\right) dr^2
\end{equation}
with the angle $\phi (t,r)$ solving the Lorentzian (Euclidean) sine--Gordon
equation,
\begin{equation}
\label{sg1}-\epsilon \frac{\partial ^2\phi }{\partial t^2}+\frac{\partial
^2\phi }{\partial r^2}=\widetilde{m}^2\sin \phi ,
\end{equation}
which determine some 2D Rimannian geometries with constant negative
curvature,
\begin{equation}
\label{curv1}\widetilde{R}=-2\widetilde{m}^2=const.
\end{equation}
The angle $\phi \left( t,r\right) $ from (\ref{soliton1}) describes an
embedding of a 2D manifold into a three--dimensional (pseudo) Euclidean
space.

The topic of construction of
 soliton solutions in gravity theories has a long
history (see Refs \cite{belinski,har,mais,neug}). For 4D vacuum Einstein
gravity the problem was tackled by investigating metrics $g_{\alpha \beta }$
of signature $(-,+,+,+)$ (in this paper permutations of sines will be
 also considered),
  with a 2+2 spacetime splitting,
\begin{equation}
\label{belmais}-ds^2=g_{ij}\left( x^i\right) dx^idx^j+h_{ab}(x^i)dy^ady^b,
\end{equation}
where $g_{ij}= diag[-f(x^i),f(x^i)]$ and $\det h_{ab}<1$ and the local
coordinates are denoted
$$
u^\alpha =\left( \{x^i=(x^1,x^2)\},\{y^a=(y^3=z,y^4)\}\right) ,
$$
or, in brief, $u=(x,y).$ We adopt the convention that
the $x$--coordinates are provided with Latin indices of type $%
i,j,k,...=1,2;$  the $y$--coordinates are with indices of type $%
a,b,c,...,=3,4$ and the 4D $u$--coordinates will be provided with
 Greek indices $\alpha , \beta , ... = 1, 2, 3, 4.$
  The meaning of coordinates (space or time like ones) will
depend on the type of construction under consideration. Belinski and
Zakharov \cite{belinski} identified $y^4$ with the time like coordinate;
Maison \cite{mais} treated $y^a$ as space variables. It was proved that the
vacuum Einstein equations, $R_{\alpha \beta }=0,$ are satisfied if the
components $h_{ab}(x^i)$ are solutions of a generalized (Euclidean)
sine--Gordon equation.

In two recent papers \cite{gegenberg} Gegenberg and Kunstatter investigated
the relationship between black holes of JT dilation gravity and solutions of
the sine--Gordon field theory. Their constructions were generalized by
Cadoni \cite{cadoni} to soliton solutions of 2D dilaton gravity models
which describes spacetimes being of non constant curvature.

In this work we explore the possibility of anholonom\-ic
ge\-ne\-ra\-li\-za\-ti\-ons of
the  Belinski--Zakharov--Mais\-on \cite{belinski,mais} soliton
constructions (\ref{belmais}) in a fashion when the coefficients of the
matrix $h_{ab}$ could depend on three variables $(x^i,z).$ The matrix $%
g_{ij}(x^i)$ will be defined by some Gegenberg--Kunstatter--Cadoni 2D
soliton--black hole solutions \cite{gegenberg,cadoni}, their conformal
transforms, or by the factor $f(x^i)$ from (\ref{belmais}) which could be
related with solitons for $h_{ab}.$ We emphasize that the resulting 4D
(pseudo) Riemannian metrics, with generic local anisotropy (in brief,
la--metrics) will be found to solve the Einstein equations with
energy--momentum tensor.

For definiteness, we consider 4D metrics parametrized by ansatzs of type
\begin{equation}
\label{ansatz}g_{\alpha \beta }=\left[
\begin{array}{cc}
g_{ij}+N_i^aN_j^bh_{ab} & N_j^eh_{ae} \\
N_i^eh_{be} & h_{ab}
\end{array}
\right]
\end{equation}
which are given with respect to a local coordinate basis $du^\alpha =\left(
dx^i,dy^a\right) $ being dual to $\partial /u^\alpha =\left( \partial
/x^i,\partial /y^a\right) .$ For simplicity, the 2D components
  $g_{ij}$ and $h_{ab}$ are considered to be some diagonal matrices
 (for two dimensions a diagonalization is always possible),
\begin{equation}
\label{hansatz}g_{ij}(x^k)=\left(
\begin{array}{cc}
g_1(x^k) & 0 \\
0 & g_2(x^k)
\end{array}
\right)
\end{equation}
and
\begin{equation}
\label{vansatz}h_{ab}(x^k,z)=\left(
\begin{array}{cc}
h_3(x^k,z) & 0 \\
0 & h_4(x^k,z)
\end{array}
\right) .
\end{equation}
The components $N_i^a=N_i^a(x^i,z)$ will be selected as to satisfy the 4D
Einstein gravitational field equations.

The metric (\ref{ansatz}) can be rewritten in a very simple form
\begin{equation}
\label{dm}g_{\alpha \beta }=\left(
\begin{array}{cc}
g_{ij}(x^k) & 0 \\
0 & h_{ab}(x^k,z)
\end{array}
\right)
\end{equation}
with respect to some 2+2 anholonomic bases (tetrads, or vierbiends) defined
\begin{eqnarray}
\delta _\alpha & = &
(\delta _i,\partial _a ) = \frac \delta {\partial u^\alpha }%
\label{dder}   \\
 & = &
 \left( \delta _i = \frac \delta {\partial x^i} =
\frac \partial {\partial x^i} -
N_i^b \left( x^j,y\right) \frac \partial {\partial y^b},
 \partial _a  = \frac \partial {\partial y^a}\right)
\nonumber
\end{eqnarray}
and
\begin{eqnarray}
\delta ^\beta & = & \left( d^i,\delta ^a \right)
  =  \delta u^\beta \label{ddif}
\\ & = &
\left( d^i = dx^i,
 \delta ^a = \delta y^a =dy^a +N_k^a \left( x^j,y^b \right) dx^k\right) .
\nonumber
\end{eqnarray}
The coefficients $N_j^a\left( u^\alpha \right) $ from (\ref{dder}) and (\ref
{ddif}) could be treated as the components of an associated nonlinear
connection, N--connection, structure (see \cite{barthel,ma,v2}; in this work
we do not consider in details the N--connection geometry).

A specific point of this paper, comparing with another soliton approaches
in gravity theories, is to show how anholonomic constructions can be
used for generation of 4D soliton--dilaton--black hole non--perturbative
structures in general relativity. This way a correspondence between
solutions of so--called locally anisotropic (super) gravity and string
theories \cite{v2} and metrics given with respect to anholonomic frames in
general relativity is derived.

Ansatzs of type (\ref{hansatz}), (\ref{vansatz}) and (\ref{dm}) can be used
for construction of an another class of solutions with generic local
anisotropy of the Einstein equations. If the gravitational field equations
are written with respect to an anholonomic basis (\ref{dder}) and/or (\ref
{ddif}), the coefficients $g_{ij},h_{ab}$ and $N_j^a$ satisfy some very
simplified systems of partial differential equations. We can construct
various classes of black hole and disk/cylinder like solutions which in the
locally isotropic limit are conformally equivalent to some well known BTZ,
Schwarzschild and/or Kerr, Weyl cylindrical and Neugebauer--Meinel disk
solutions. In general relativity there are admitted singular (in a point, on
unclosed infinite lines or on closed curves such as ellipses, circles)
solutions with horizons being described by hypersurface equations for
rotation ellipsoid, torus, ellipses and so on. A physical treatment of such
nonlinear configurations is to consider values like anisotropic mass,
oscillation of horizons, variable  interaction constants and gravitational
 non--linear self polarizations.

\subsection{Outline}

The paper is organized as follow:

Section II reviews the geometry of anholonomic frames on (pseudo) Riemannian
spaces and associated nonlinear connection structures. There are defined the
basic geometric objects and written the Einstein equations with respect to
anholonomic frames split by nonlinear connections.

In Section III, there are considered the general properties and reductions of
basic geometric objects and field equations for 4D metrics constructed as
nonlinear superpositions of 2D horizontal (with respect to a nonlinear
connection structure) metrics, depending on two horizontal coordinates, and
of 2D vertical coordinates depending on three (two horizontal plus one
vertical) coordinates. It is given a classification of such 4D metrics
depending on signatures of 2D metrics and resulting 4D metrics.

In Section IV, we prove that the Einstein equations admit soliton like 2D
(both for horizontal and vertical components of 4D metrics) and 3D (for
vertical components of 4D metrics) solutions. There are examined some
classes of integral varieties for the Einstein equations admitting
non--perturbative structures generated, for instance, by 2D and 3D
sine--Gordon and Kadomtsev--Petviashvili equations. Some exact solutions for
locally anisotropic deformations of the sine--Gordon systems are constructed.

Section V describes an effective locally anisotropic soliton--dilaton field
theory and contains a topological analysis of such models.

Section VI elucidates the interconnection of locally an\-isot\-rop\-ic 2D
soliton and black hole solutions. Nonlinear superpositions to 4D are
considered.

In Section VII, we construct 3D black hole solutions with generic local
anisotropic. As  some simplest examples there
are taken configurations when the horizon is
parametrized by an ellipse and the possibility of oscillation in time of such
horizons is shown.

Section VIII is devoted to the physics of 4D black hole and disk/cylinder
solutions with generic local anisotropy. There are analyzed the general
properties of metrics describing such solutions and discussed the question
of their physical treatment. The construction of singular
solutions with various type of horizon
 hypersurfaces is performed by considering
correspondingly the rotation ellipsoid, epllipsoidal cylinder, torus,
bipolar and another systems of coordinates. It is shown that in the locally
isotropic limit such solutions could be equivalent to some conformal
transforms of some static or rotating configurations like for the
Schwarzschild--Kerr, Weyl cylindrical and Neugebauer--Meinel disk solutions.

In Section IX, some additional examples of local\-ly an\-isot\-rop\-ic
soliton--dilaton--black hole solutions are given. It is illustrated how
 in general relativity we can
 construct two soliton  non--perturbative structures,  proved that
 nonlinear connections and non--diagonal energy--momentum tensor
 components can induce Kadomtsev--Petviashvily soliton like solutions
 and there are considered new types   of two and three coordinate
  soliton--dilaton  vacuum gravitational configurations.

Finally, in Section X, we discuss our results and present conclusions.

\section{ Anholonomic Frames on (Pseudo) Riemannian Spaces}

We outline the geometric background on anholonomic frames modelling 2D local
anisotropies (la) in 4D curved spaces \cite{v2} (see Refs. \cite{haw} and
\cite{ma} for details on spacetime differential geometry and N--connection
structures in vector bundle spaces). We note that a frame anholonomy induces
a corresponding local anisotropy. Spacetimes enabled with anholonomic frame
(and associated N--connection) structures are also called locally
anisotropic spacetimes, in brief la--spacetimes.

In this paper spacetimes are modelled as smooth (i.e class $C^\infty )$ 4D
(pseudo) Riemannian manifolds $V^{(3+1)}$ being Hausdorff, paracompact and
connected and provided with corresponding geometric structures of symmetric
metric $g_{\alpha \beta }$ of signature $\left( -,+,+,+\right) $ and of
linear, in general nonsymmetric, connection $\Gamma _{~\beta \gamma }^\alpha
$ defining the covariant derivation $D_\alpha $ satisfying the metricity
conditions $D_\alpha g_{\beta \gamma }=0.$ The indices are given with
respect to a tetradic (frame) vector field $\delta ^\alpha =(\delta
^i,\delta ^a)$ and its dual $\delta _\alpha =(\delta _i,\delta _a).$

A frame (local basis) structure $\delta _\alpha $ (\ref{ddif}) on $V^{(3+1)}$
is characterized by its anholonomy coefficients $w_{~\beta \gamma }^\alpha $
defined from relations
\begin{equation}
\label{anholon}\delta _\alpha \delta _\beta -\delta _\beta \delta _\alpha
=w_{~\alpha \beta }^\gamma \delta _\gamma .
\end{equation}

The elongation (by N--coefficients) of partial derivatives in the locally
adapted partial derivatives (\ref{dder}) reflects the fact that on the
(pseudo) Riemannian spacetime $V^{(3+1)}$ it is modelled a generic local
anisotropy characterized by anholonomy relations (\ref{anholon}) when the
anholonomy coefficients are computed as follows
\begin{eqnarray}
w_{~ij}^k & = & 0,w_{~aj}^k=0,w_{~ia}^k=0,w_{~ab}^k=0,w_{~ab}^c=0,
\nonumber\\
w_{~ij}^a & = &
-\Omega _{ij}^a,w_{~aj}^b=-\partial _aN_i^b,w_{~ia}^b=\partial _aN_i^b,
\nonumber
\end{eqnarray}
where%
$$
\Omega _{ij}^a=\partial _iN_j^a-\partial _jN_i^a+N_i^b\partial
_bN_j^a-N_j^b\partial _bN_i^a
$$
defines the coefficients of N--connection curvature, in brief, N--curvature.
On (pseudo) Riemannian spaces this is a characteristic of a chosen
anholonomic system of reference.

A 2+2 anholonomic structure distinguishes (d) the geometrical objects into
horizontal (h) and vertical (v) components. Such objects are briefly called
d--tensors, d--metrics and/or d--connections. Their components are defined
with respect to a la--basis of type (\ref{dder}), its dual (\ref{ddif}), or
their tensor products (d--linear or d--affine transforms of such frames
could also be considered). For instance, a covariant and contravariant
d--tensor $Z,$ is expressed
\begin{eqnarray}
Z &= & Z_{~\beta }^\alpha \delta _\alpha \otimes \delta ^\beta  \nonumber \\
 {} &= & Z_{~j}^i\delta _i\otimes d^j+Z_{~a}^i\delta _i\otimes \delta
^a+Z_{~j}^b\partial _b\otimes d^j+Z_{~a}^b\partial _b\otimes \delta ^a.
\nonumber
\end{eqnarray}

A linear d--connection $D$ on la--space $V^{(3+1)}{\cal ,}$
$$
D_{\delta _\gamma }\delta _\beta =\Gamma _{~\beta \gamma }^\alpha \left(
x,y\right) \delta _\alpha ,
$$
is parametrized by non--trivial h--v--components,
\begin{equation}
\label{dcon}\Gamma _{~\beta \gamma }^\alpha =\left(
L_{~jk}^i,L_{~bk}^a,C_{~jc}^i,C_{~bc}^a\right) .
\end{equation}

A metric on $V^{(3+1)}$ with 2+2 block coefficients (\ref{dm}) is written in
distinguished form, as a metric d--tensor (in brief, d--metric), with respect
 a la--base (\ref{ddif})
\begin{eqnarray}
\delta s^2 & = & g_{\alpha \beta }\left( u\right)
  \delta ^\alpha \otimes \delta^\beta \label{dmetric} \\
{} & = & g_{ij}(x,y)dx^idx^j+h_{ab}(x,y)\delta y^a\delta y^b. \nonumber
\end{eqnarray}

Some d--connection and d--metric structures are compatible if there are
satisfied the conditions
$$
D_\alpha g_{\beta \gamma }=0.
$$
For instance, a canonical compatible d--connection
$$
^c\Gamma _{~\beta \gamma }^\alpha =\left(
^cL_{~jk}^i,^cL_{~bk}^a,^cC_{~jc}^i,^cC_{~bc}^a\right)
$$
is defined by the coefficients of d--metric (\ref{dmetric}), $g_{ij}\left(
x,y\right) $ and $h_{ab}\left( x,y\right) ,$ and by the N--coefficients,%
\begin{eqnarray}
^cL_{~jk}^i & = & \frac 12g^{in}\left( \delta _kg_{nj}+\delta _jg_{nk}-\delta
_ng_{jk}\right) , \label{cdcon} \\
^cL_{~bk}^a & = & \partial _bN_k^a+\frac 12h^{ac}\left( \delta
_kh_{bc}-h_{dc}\partial _bN_i^d-h_{db}\partial _cN_i^d\right) ,
\nonumber \\
^cC_{~jc}^i & = & \frac 12g^{ik}\partial _cg_{jk}, \nonumber \\
^cC_{~bc}^a & = & \frac 12h^{ad}\left( \partial _ch_{db}+\partial
_bh_{dc}-\partial _dh_{bc}\right)  \nonumber
\end{eqnarray}
The coefficients of the canonical d--connection generalize for
la--spacetimes the well known Cristoffel symbols.

For a d--connection (\ref{dcon}) the components of torsion,
\begin{eqnarray}
&T\left( \delta _\gamma ,\delta _\beta \right) &=T_{~\beta \gamma }^\alpha
\delta _\alpha ,  \nonumber \\
&T_{~\beta \gamma }^\alpha &=
\Gamma _{~\beta \gamma }^\alpha -\Gamma _{~\gamma
\beta }^\alpha +w_{~\beta \gamma }^\alpha \nonumber
\end{eqnarray}
are expressed via d--torsions
\begin{eqnarray}
T_{.jk}^i & = & T_{jk}^i=L_{jk}^i-L_{kj}^i,\quad
T_{ja}^i=C_{.ja}^i,T_{aj}^i=-C_{ja}^i, \nonumber \\
T_{.ja}^i & = & 0,\quad T_{.bc}^a=S_{.bc}^a=C_{bc}^a-C_{cb}^a,
\label{dtors} \\
T_{.ij}^a & = &
-\Omega _{ij}^a,\quad T_{.bi}^a= \partial _b  N_i^a
-L_{.bj}^a,\quad T_{.ib}^a=-T_{.bi}^a. \nonumber
\end{eqnarray}
We note that for symmetric linear connections the d--torsion is induced as a
pure anholonomic effect.

In a similar manner, putting non--vanishing coefficients (\ref{dcon}) into
the formula for curvature
\begin{eqnarray}
&R\left( \delta _\tau ,\delta _\gamma \right) \delta _\beta &=
R_{\beta ~\gamma\tau }^{~\alpha }\delta _\alpha , \nonumber \\
&R_{\beta ~\gamma \tau }^{~\alpha } & =  \delta _\tau
\Gamma _{~\beta \gamma
}^\alpha -\delta _\gamma \Gamma _{~\beta \delta }^\alpha +
\nonumber \\
& & \Gamma _{~\beta \gamma }^\varphi \Gamma _{~\varphi \tau }^\alpha
-\Gamma
_{~\beta \tau }^\varphi \Gamma _{~\varphi \gamma }^\alpha +
\Gamma _{~\beta
\varphi }^\alpha w_{~\gamma \tau }^\varphi,  \nonumber
\end{eqnarray}
we can compute the components of d--curvatures
\begin{eqnarray}
R_{h.jk}^{.i} & = & \delta _kL_{.hj}^i-\delta_jL_{.hk}^i \nonumber \\
 & & +  L_{.hj}^mL_{mk}^i-L_{.hk}^mL_{mj}^i-C_{.ha}^i\Omega _{.jk}^a,
\nonumber \\
R_{b.jk}^{.a} & = & \delta _kL_{.bj}^a-\delta_jL_{.bk}^a \nonumber \\
 & & +  L_{.bj}^cL_{.ck}^a-L_{.bk}^cL_{.cj}^a-C_{.bc}^a\Omega _{.jk}^c,
\nonumber \\
P_{j.ka}^{.i} & = & \partial _kL_{.jk}^i +C_{.jb}^iT_{.ka}^b \nonumber \\
 & & -  ( \partial _kC_{.ja}^i+L_{.lk}^iC_{.ja}^l -
L_{.jk}^lC_{.la}^i-L_{.ak}^cC_{.jc}^i ), \nonumber \\
P_{b.ka}^{.c} & = & \partial _aL_{.bk}^c +C_{.bd}^cT_{.ka}^d \nonumber \\
 & & - ( \partial _kC_{.ba}^c+L_{.dk}^{c\,}C_{.ba}^d
- L_{.bk}^dC_{.da}^c-L_{.ak}^dC_{.bd}^c ) \nonumber \\
S_{j.bc}^{.i} & = & \partial _cC_{.jb}^i-\partial _bC_{.jc}^i
 +  C_{.jb}^hC_{.hc}^i-C_{.jc}^hC_{hb}^i, \nonumber \\
S_{b.cd}^{.a} & = &\partial _dC_{.bc}^a-\partial
_cC_{.bd}^a+C_{.bc}^eC_{.ed}^a-C_{.bd}^eC_{.ec}^a. \nonumber
\end{eqnarray}

The Ricci tensor
$$
R_{\beta \gamma }=R_{\beta ~\gamma \alpha }^{~\alpha }
$$
has the d--components
\begin{eqnarray}
R_{ij} & = & R_{i.jk}^{.k},\quad
 R_{ia}=-^2P_{ia}=-P_{i.ka}^{.k},\label{dricci} \\
R_{ai} &= & ^1P_{ai}=P_{a.ib}^{.b},\quad R_{ab}=S_{a.bc}^{.c}. \nonumber
\end{eqnarray}
We point out that because, in general, $^1P_{ai}\neq ~^2P_{ia},$ the Ricci
d-tensor is non symmetric.

Having defined a d-metric of type (\ref{dmetric}) in $V^{(3+1)}$ we can
compute the scalar curvature
$$
\overleftarrow{R}=g^{\beta \gamma }R_{\beta \gamma }.
$$
of a d-connection $D,$%
\begin{equation}
\label{dscalar}{\overleftarrow{R}}=G^{\alpha \beta }R_{\alpha \beta }=%
\widehat{R}+S,
\end{equation}
where $\widehat{R}=g^{ij}R_{ij}$ and $S=h^{ab}S_{ab}.$

Now, by introducing the values (\ref{dricci}) and (\ref{dscalar}) into the
Einstein's equations
$$
R_{\beta \gamma }-\frac 12g_{\beta \gamma }\overleftarrow{R}=k\Upsilon
_{\beta \gamma },
$$
we can write down the system of field equations for la--gravity with
prescribed anholonomic (N--connection) structure \cite{ma}:%
\begin{eqnarray}
R_{ij}-\frac 12\left( \widehat{R}+S\right) g_{ij} & = &
k\Upsilon _{ij}, \label{einsteq2} \\
S_{ab}-\frac 12\left( \widehat{R}+S\right) h_{ab} & = & k\Upsilon _{ab},
 \nonumber \\
^1P_{ai} & = & k\Upsilon _{ai}, \nonumber \\
^2P_{ia} & = & -k\Upsilon _{ia}, \nonumber
\end{eqnarray}
where $\Upsilon _{ij},\Upsilon _{ab},\Upsilon _{ai}$ and $\Upsilon _{ia}$
are the components of the energy--momentum d--tensor field $\Upsilon _{\beta
\gamma }$ (which includes possible cosmological constants,
contributions of anholonomy d--torsions (\ref{dtors}) and matter) and $k$ is
the coupling constant

\section{4D Anholonomic Superpositions of 2D D--Metrics}

Let us consider a 4D spacetime $V^{(3+1)}$ provided with a d--metric (\ref
{dmetric}) when $g_i = g_i (x^k)$ and $h_a = h_a (x^k, z)$  for
$y^a = (z, y^4).$
 The N--connection coefficients are
restricted to be some functions on three coordinates $(x^i,z),$%
\begin{eqnarray}
N_1^3&=&q_1(x^i,z),\ N_2^3=q_2(x^i,z), \label{ncoef} \\
N_1^4&=&n_1(x^i,z),\ N_2^4=n_2(x^i,z). \nonumber
\end{eqnarray}
For simplicity, we shall use brief denotations of partial derivatives, like $%
\dot a$$=\partial a/\partial x^1,a^{\prime }=\partial a/\partial x^2,$ $%
a^{*}=\partial a/\partial z$ $\dot a^{\prime }$$=\partial ^2a/\partial
x^1\partial x^2,$ $a^{**}=\partial ^2a/\partial z\partial z.$

The non--trivial components of the Ricci d--tensor (\ref{dricci}),
 for the mentioned type of d--metrics depending on three variables, are%
\begin{eqnarray}
&R_1^1&=R_2^2=\frac 1{2g_1g_2} \times \label{ricci1}\\
&{} &
[-(g_1^{^{\prime \prime }}+{\ddot g}_2)+\frac 1{2g_2}\left( {\dot g}%
_2^2+g_1^{\prime }g_2^{\prime }\right) +\frac 1{2g_1}\left( g_1^{\prime \ 2}+%
\dot g_1\dot g_2\right) ]; \nonumber
\end{eqnarray}
\begin{equation}
\label{ricci2}S_3^3=S_4^4=\frac 1{h_3h_4}[-h_4^{**}+\frac 1{2h_4}(h_4^{*})^2+%
\frac 1{2h_3}h_3^{*}h_4^{*}];
\end{equation}
\begin{eqnarray}
P_{31}&=&\frac{q_1}2[\left( \frac{h_3^{*}}{h_3}\right) ^2-
\frac{h_3^{**}}{h_3}+%
\frac{h_4^{*}}{2h_4^{\ 2}}-\frac{h_3^{*}h_4^{*}}{2h_3h_4}] \label{ricci3}
\\
&{} &
+\frac 1{2h_4}[\frac{\dot h_4}{2h_4}h_4^{*}-\dot h_4^{*}+ %
\frac{\dot h_3}{2h_3}h_4^{*}],  \nonumber \\
P_{32}&=&
\frac{q_2}2[\left( \frac{h_3^{*}}{h_3}\right) ^2-\frac{h_3^{**}}{h_3}+%
\frac{h_4^{*}}{2h_4^{\ 2}}-\frac{h_3^{*}h_4^{*}}{2h_3h_4}] \nonumber \\
&{} &
+\frac 1{2h_4}[\frac{h_4^{\prime }}{2h_4}h_4^{*}-h_4^{\prime \ *}+ %
\frac{h_3^{\prime }}{2h_3}h_4^{*}];  \nonumber
\end{eqnarray}
\begin{eqnarray}
 P_{41}&=&-\frac{h_4}{2h_3}n_1^{**}, \label{ricci4} \\
 P_{42}&=&-\frac{h_4}{2h_3}n_2^{**}. \nonumber
\end{eqnarray}

The curvature scalar $\overleftarrow{R}$ (\ref{dscalar}) is defined by two
non-trivial components $\widehat{R}=2R_1^1$ and $S=2S_3^3.$

The system of Einstein equations (\ref{einsteq2}) transforms into
\begin{eqnarray}
R_1^1&=&-\kappa \Upsilon _3^3=-\kappa \Upsilon _4^4,
\label{einsteq3a} \\
S_3^3&=&-\kappa \Upsilon _1^1=-\kappa \Upsilon _2^2, \label{einsteq3b}\\
P_{3i}&=& \kappa \Upsilon _{3i}, \label{einsteq3c} \\
P_{4i}&=& \kappa \Upsilon _{4i}, \label{einsteq3d}
\end{eqnarray}
where the values of $R_1^1,S_3^3,P_{ai},$ are taken respectively from (\ref
{ricci1}), (\ref{ricci2}), (\ref{ricci3}), (\ref{ricci4}).

By using the equations (\ref{einsteq3c}) and (\ref{einsteq3d}) we can define
the N--coefficients (\ref{ncoef}), $q_i(x^k,z)$ and $n_i(x^k,z),$ if the
functions $h_i(x^k,z)$ are known as solutions of the equations (\ref
{einsteq3b}).

Now, we discuss the question on possible signatures of generated 4D metrics.
There are three classes of la--solutions:

\begin{enumerate}
\item  The horizontal d--metric is fixed to be of Lorentzian signature, $%
sign\left( g_{ij}\right) =(-,+),$ the vertical one is of Euclidean
signature, $sign\left( h_{ab}\right) =(+,+),$ and the resulting 4D metric $%
g_{\alpha \beta }$ will be considered of signature $(-,+,+,+).$ The local
coordinates are chosen $u^\alpha =(x^1=t,x^2,y^3=z,y^4),$ where $t$ is the
time like coordinate and the d--metrics are parametrized
\begin{equation}
\label{dm2drmh}g_{ij}\left( t,x^1\right) =\left(
\begin{array}{cc}
g_1=-\exp a_1 & 0 \\
0 & g_2=\exp a_2
\end{array}
\right)
\end{equation}
and
\begin{equation}
\label{dm2drmv}h_{ab}\left( t,x^1,z\right) =\left(
\begin{array}{cc}
h_3=\exp b_3 & 0 \\
0 & h_4=\exp b_4
\end{array}
\right) .
\end{equation}
The energy--momentum d--tensor for the Einstein equations (\ref{einsteq2})
 could be considered in diagonal form
\begin{equation}
\label{dem}\Upsilon _\beta ^\alpha =diag[-\varepsilon ,p_2,p_3,p_4]
\end{equation}
if the N--coefficients $N_i^a(t,x^1,z)$ are chosen to make zero the
non--diagonal components of the Ricci d--tensor (see (\ref{ricci3}) and (\ref
{ricci4})). Here we note that on la--spacetimes, with respect to anholonomic
frames, there are possible nonzero values of pressure, $p\neq 0,$ even $%
\varepsilon =0.$

\item  The horizontal d--metric is fixed to be of Euclidean signature, $%
sign\left( g_{ij}\right) =(+,+),$ the vertical one is of Lorentzian
signature, $sign\left( h_{ab}\right) =(+,-)$ and the resulting 4D metric $%
g_{\alpha \beta }$ will be considered to be a static one with signature $%
(+,+,+,-).$ The local coordinates are chosen $u^\alpha
=(x^1,x^2,y^3=z,y^4=t)$ and the d--metrics are parametrized
\begin{equation}
\label{dm2drmh1}g_{ij}\left( x^k\right) =\left(
\begin{array}{cc}
g_1=\exp a_1 & 0 \\
0 & g_2=\exp a_2
\end{array}
\right)
\end{equation}
and
\begin{equation}
\label{dm2drmv1}h_{ab}\left( x^k,z\right) =\left(
\begin{array}{cc}
h_3=\exp b_3 & 0 \\
0 & h_4=-\exp b_4
\end{array}
\right) .
\end{equation}
The energy--momentum d--tensor for the Einstein equations (\ref{einsteq2})
could be considered in diagonal form
\begin{equation}
\label{dem1}\Upsilon _\beta ^\alpha =diag[p_1,p_2,p_3,-\varepsilon ]
\end{equation}
if the coefficients $N_i^a(x^i,z)$ are chosen to diagonalize the Ricci
d--tensor (when (\ref{ricci3}) and (\ref{ricci4}) are zero).

\item  The horizontal d--metric is fixed to be of Euclidean signature, $%
sign\left( g_{ij}\right) =(+,+),$ the vertical one is of Lorentzian
signature, $sign\left( h_{ab}\right) =(-,+).$ The local coordinates are
chosen $u^\alpha =(x^1,x^2,y^3=z=t,y^4)$ and the d--metrics are parametrized
\begin{equation}
\label{dm2drmh2}g_{ij}\left( x^k\right) =\left(
\begin{array}{cc}
g_1=\exp a_1 & 0 \\
0 & g_2=\exp a_2
\end{array}
\right)
\end{equation}
and
\begin{equation}
\label{dm2drmv2}h_{ab}\left( x^k,t\right) =\left(
\begin{array}{cc}
h_3=-\exp b_3 & 0 \\
0 & h_4=\exp b_4
\end{array}
\right) .
\end{equation}
The energy--momentum d--tensor for the Einstein equations (\ref{einsteq2})
is considered in diagonal form
\begin{equation}
\label{dem2}\Upsilon _\beta ^\alpha =diag[p_1,p_2,-\varepsilon ,p_4]
\end{equation}
if the N--coefficients $N_j^a(x^i,t)$ make zero the non--diagonal components
of the Ricci d--tensor (with vanishing (\ref{ricci3}) and (\ref{ricci4})).
\end{enumerate}

The following Sections are devoted to a general study and explicit
constructions of 4D solutions of the Einstein equations via type 1--3
nonlinear superpositions of 2D soliton--dilaton--black hole d--metrics.

\section{Locally Anisotropic Soliton Like Equations}

We have found in the last Section that the vertical component of
energy--momentum d--tensor is the non--vacuum source of the horizontal
components of a d--metric (see the equations (\ref{einsteq3a})) and,
inversely, following (\ref{einsteq3b}), one could conclude that the horizontal
component of energy--momentum d--tensor is the non--vacuum source of the
vertical components of a d--metric. The horizontal and vertical components
of the distinguished Einstein equations (\ref{einsteq2}) are rather
different by structures and this is taken into account by
 choosing the metric ansatz (\ref{ansatz}) which gives rise in a very
 simplified form of partial differential equations.
 If the 2D h--metric depends on two variables, $%
g_{ij}=g_{ij}(x^k),$ with the diagonal components satisfying a second order
partial differential equation with respect to 'dot' and 'prime' derivatives,
the 2D v--metric could be on three variables, $h_{ab}=h_{ab}(x^k,z),$ with
diagonal components satisfying a second order partial derivation with
respect to 'star' derivatives. The purpose of this Section is to prove that
both type of Einstein h-- and v--equations admit soliton like solutions.

\subsection{Horizontal La--Deformed Sine--Gordon Equations}

Let us parametrize the horizontal part of d--metric (h--metric) as
\begin{eqnarray}
g_1 &=& \epsilon \sin ^2\left[ v\left( x^i\right) /2\right] ,\epsilon =\pm 1,
 \nonumber \\
 g_2 &=& \cos ^2\left[ v\left( x^i\right) /2\right] . \nonumber
\end{eqnarray}
The non--trivial component of the
Ricci d--tensor (\ref{ricci1}) is written
\begin{equation}
\label{riccihs}\epsilon R_1^1=\frac 1{\sin v}\left( \ddot v-\epsilon
v^{\prime \prime }\right) +\rho \left( x^i\right)
\end{equation}
where
\begin{equation}
\label{rho}\rho \left( x^i\right) =\rho \left( v,\dot v,v^{\prime }\right) =%
\frac{\cos v-1}{\sin ^2v}\left( \dot v^2-\epsilon v^{\prime \ 2}\right) .
\end{equation}

The horizontal Einstein equations (\ref{einsteq3a}) are
\begin{equation}
\label{sgla}\left( \ddot v-\epsilon v^{\prime \prime }\right) +\frac{\cos v-1%
}{\sin v}\left( \dot v^2-\epsilon v^{\prime \ 2}\right) =\epsilon \kappa
\Upsilon _3^3\ \sin v,
\end{equation}
being compatible for matter states when $\Upsilon _3^3=\Upsilon _4^4.$ For
simplicity, we consider the case of constant energy density or pressure
 (depending on the type of fixed signature),
$\Upsilon _3^3=\Upsilon _3=const.$
The equation (\ref{sgla}) defines some
components of a 4D metric (\ref{dm}) and is defined by a locally anisotropic
deformation of the Euclidean (for $\epsilon =-1),$ or Lorentzian (for $%
\epsilon =-1),$ sine--Gordon equation (\ref{sg1}), which are related with 2D
(pseudo) Riemannian metrics (\ref{soliton1}) and constant curvatures (\ref
{curv1}).

The first term from (\ref{riccihs}) transforms into a negative constant, $-%
\widetilde{m}^2,$ if the function $v\left( x^i\right) $ is
 chosen to be a soliton type one
which solves the sine--Gordon equation (\ref{sg1}). The physical
interpretation of terms depends of the type of the solutions we try to
construct (see below). We note that if $\epsilon =-1,$ the second term, $%
\rho (x^i),$ from (\ref{rho}) is connected with the energy density of the
soliton wave%
$$
H=\frac 12\left( \dot v^2+v^{\prime \ 2}\right) +1-\cos v.
$$

The aim of this Subsection is to investigate some basic properties of the
locally anisotropic sine--Gordon equation, which describes a 2D horizontal
soliton--dilaton system (see the next Section) induced anholonomically via a
source $\Upsilon _3$ in the vertical subspace.

\subsubsection{Lorentzian la--soliton systems of Class 1}

We consider the Class 1 of Lorentzian h--metrics (\ref{dm2drmh}) which in
local coordinates $x^i=(t,r),$ $t$ is a time like coordinate, and for $%
\epsilon =-1$ and $\Upsilon _3=p_3+(1/2\kappa )\lambda ,$ where $p_3$ is the
anisotropic pressure in the $z$--direction and $\lambda $ is the cosmological
constant, are defined by the equation
\begin{equation}
\label{sgla1}\left( \ddot v+v^{\prime \prime }\right) +\frac{\cos v-1}{\sin v%
}\left( \dot v^2+v^{\prime \ 2}\right) =-(\kappa p_3+\frac \lambda 2)\sin v ,
\end{equation}
where $\dot v=\partial v/\partial t$ and $v^{\prime }=\partial v/\partial r.$

For $\cos v\simeq 1,$ by neglecting quadratic terms $v^2,$ we can approximate
the solution of (\ref{sgla1}) by a solution of the Euclidean 2D sine--Gordon
equation (\ref{sg1}) with the constant $\widetilde{m}^2=-(\kappa p_3+\frac
\lambda 2)$. If we wont to treat $\widetilde{m}^2$ as a mass like constant,
we must suppose that matter is in a state for which $(\kappa p_3+\frac
\lambda 2)<0.$

\subsubsection{Euclidean la--soliton systems of Class 2}

This type of h--metrics (\ref{dm2drmh1}), of Class 2 from the previous
 Section, for $\epsilon =1,$ $\Upsilon _3=p_3+(1/2\kappa )\lambda $ and both
space like local coordinates $x^i,$ is given by the following from (\ref
{sgla}) equations
\begin{equation}
\label{sgla2}\left( \ddot v-v^{\prime \prime }\right) +\frac{\cos v-1}{\sin v%
}\left( \dot v^2-v^{\prime \ 2}\right) =(\kappa p_3+\frac \lambda 2)\ \sin
v,
\end{equation}
where $\dot v=\partial v/\partial x^1$ and $v^{\prime }=\partial v/\partial
x^2,$ which for $\cos v\simeq 1$ has solutions approximated by the
Lorentzian 2D sine--Gordon equation with $\widetilde{m}^2=(\kappa p_3+\frac
\lambda 2).$

\subsubsection{Euclidean la--soliton systems of Class 3}

In this case the h--metrics (\ref{dm2drmh2}) of Class 3 from the previous
Section, for $\epsilon =1,$ $\Upsilon _3=-\varepsilon +(1/2\kappa )\lambda $
and both space like local coordinates $x^i,$ is defined by equations
\begin{equation}
\label{sgla3}\left( \ddot v-v^{\prime \prime }\right) +\frac{\cos v-1}{\sin v%
}\left( \dot v^2-v^{\prime \ 2}\right) =(-\kappa \varepsilon +\frac \lambda 2%
)\ \sin v,
\end{equation}
where $\dot v=\partial v/\partial x^1$ and $v^{\prime }=\partial v/\partial
x^2,$ which for $\cos v\simeq 1$ has solutions approximated by the
Lorentzian 2D sine--Gordon equation with $\widetilde{m}^2=(-\kappa \epsilon +%
\frac \lambda 2).$

\subsubsection{A static one dimensional exact solution}

The la--deformed sine--Gordon equations can be integrated exactly for static
configurations. If $v=v_s(x^2=x),v^{\prime }=dv_s/dx$ the equation (\ref
{sgla}) transforms into
\begin{equation}
\label{sglast}\frac{d^2v_s}{dx^2}+\frac{\cos v_s-1}{\sin v_s}\left( \frac{%
dv_s}{dx}\right) ^2+\kappa \Upsilon _3\ \sin v_s=0.
\end{equation}
which does not depend on values of $\epsilon =\pm 1.$ Introducing a new
variable $y(v_s)=\left( dv_s/dx\right) ^2$ we get a linear first order
differential equation%
$$
\frac{dy}{dv_s}+2\frac{\cos v_s-1}{\sin v_s}y+2\kappa \Upsilon _3\ \sin
v_s=0
$$
which is solved by applying standard methods \cite{kamke}.

\subsection{Vertical Einstein Equations and Possible Soliton Like Solutions
}

The basic equation
\begin{eqnarray}
\label{heq}
&{}&
\frac{\partial ^2h_4}{\partial z^2} - %
\frac 1{2h_4}\left( \frac{\partial h_4}{\partial z}\right) ^2  \\ %
&{}& -\frac 1{2h_3}\left( \frac{\partial h_3}{\partial z}\right)
\left( \frac{\partial h_4}{\partial z}\right) -
\frac \kappa 2\Upsilon _1h_3h_4=0 \nonumber %
\end{eqnarray}
(here we write down the partial derivatives on $z$ in explicit form) follows
from (\ref{ricci2}) and (\ref{einsteq3b}) and relates some first and second
order partial on $z$ derivatives of diagonal components $h_a(x^i,z)$ of a
v--metric with a source $\kappa \Upsilon _1(x^i,z)=\kappa \Upsilon
_1^1=\kappa \Upsilon _2^2$ in the h--subspace. We can consider as unknown
the function $h_3(x^i,z)$ (or, inversely, $h_4(x^i,z))$ for some compatible
values of $h_4(x^i,z)$ (or $h_3(x^i,z))$ and source $\Upsilon _1(x^i,z).$

The structure of equation (\ref{heq}) differs substantially from the
horizontal one (\ref{einsteq3a}), or (\ref{sgla}). In this Subsection we
analyze some soliton type integral varieties which solve the partial
differential equation (\ref{heq}).

\subsubsection{Belinski--Zakharov--Maison locally isotropic limits}

In the vacuum case $\Upsilon _1^1\equiv 0$ and arbitrary two functions
depending only on variables $x^i$ are admitted as solutions of (\ref{heq}).
The N--coefficients $q_i$ and $n_i$ became zero in consequence of (\ref
{einsteq3c}), (\ref{einsteq3d}) and (\ref{ricci3}), (\ref{ricci4}). So, in
the locally isotropic vacuum limit the 4D metrics (\ref{ansatz}) will
transform into a soliton vacuum solution of Einstein equations if, for
instance, we require that $h_a(x^i)$ are the components of the diagonalized
matrix for the Belinski--Zakharov \cite{belinski} or Maison \cite{mais}
gravitational solitons and the h--metric transforms is defined by a
conformal factor $f(x^i)$ being compatible with the v--metric. Of course,
instead of soliton ones we can choose another class of vacuum solutions
depending on variables $x^i$ to be the locally isotropic limit of
anholonomic gravitational systems.

\subsubsection{Kadomtsev--Petviashvili v--solitons}

By straightforward verification we conclude that the v--metric component $%
h_4(x^i,z)$ could be a solution of Kadomtsev--Petviashvili (KdP) equation
\cite{kad} (the first methods of integration of 2+1 dimensional soliton
equations where developed by Dryuma \cite{dryuma} and Zakharov and Shabat
\cite{zakhsh})
\begin{equation}
\label{kdp}h_4^{**}+\epsilon \left( \dot h_4+6h_4h_4^{\prime }+h_4^{\prime
\prime \prime }\right) ^{\prime }=0,\epsilon =\pm 1,
\end{equation}
if the component $h_3(x^i,z)$ satisfies the Bernoulli equations \cite{kamke}
\begin{equation}
\label{bern1}h_3^{*}+Y\left( x^i,z\right) (h_3)^2+F_\epsilon \left(
x^i,z\right) h_3=0,
\end{equation}
where, for $h_4^{*}\neq 0,$%
\begin{equation}
\label{sourse1}Y\left( x^i,z\right) =\kappa \Upsilon _1^1\frac{h_4}{h_4^{*}}%
,
\end{equation}
and
$$
F_\epsilon \left( x^i,z\right) =\frac{h_4^{*}}{h_4}+\frac{2\epsilon }{h_4^{*}%
}\left( \dot h_4+6h_4h_4^{\prime }+h_4^{\prime \prime \prime }\right)
^{\prime }.
$$
The three dimensional integral variety of (\ref{bern1}) is defined by
formulas
$$
h_3^{-1}\left( x^i,z\right) =h_{3(x)}^{-1}\left( x^i\right) E_\epsilon
\left( x^i,z\right) \times \int \frac{Y\left( x^i,z\right) }{E_\epsilon
\left( x^i,z\right) }dz,
$$
where
$$
E_\epsilon \left( x^i,z\right) =\exp \int F_\epsilon \left( x^i,z\right) dz
$$
and $h_{3(x)}\left( x^i\right) $ is a nonvanishing function.

In the vacuum case $Y\left( x^i,z\right) =0$ and we can write the integral
variety of (\ref{bern1})
$$
h_3^{(vac)}\left( x^i,z\right) =h_{3(x)}^{(vac)}\left( x^i\right) \exp
\left[ -\int F_\epsilon \left( x^i,z\right) dz\right] .
$$

We conclude that a solution of KdP equation (\ref{bern1}) could generate a
non--perturbative component $h_4(x^i,z)$ of a diagonal h--metric if the
second component $h_3\left( x^i,z\right) $ is a solution of Bernoulli
equations (\ref{bern1}) with coefficients determined both by $h_4$ and its
partial derivatives and by the $\Upsilon _1^1$ component of the
energy--momentum d--tensor (see (\ref{sourse1})). In the non--vacuum case
the parameters of (2+1) dimensional KdV solitons are connected with
parameters defining the interactions with matter fields and/or by a
cosmological constant. The further developments in this direction
consist in construction of self--consistent (2+1) KdV soliton solutions
induced by some soliton configurations from energy--momentum tensor in
hydrodynamical (plasma) approximations.

\subsubsection{(2+1) sine--Gordon v--solitons}

In a symilar manner as in previous paragraph we can prove that solutions $%
h_4(x^i,z)$ of (2+1) sine--Gordon equation (see, for instance, \cite
{har,lieb,whith})%
$$
h_4^{**}+h_4^{^{\prime \prime }}-\ddot h_4=\sin (h_4)
$$
also induce solutions for $h_3\left( x^i,z\right) $ following from the
Bernoulli equation%
$$
h_3^{*}+\kappa \Upsilon _1(x^i,z)\frac{h_4}{h_4^{*}}(h_3)^2+F\left(
x^i,z\right) h_3=0,h_4^{*}\neq 0,
$$
where
$$
F\left( x^i,z\right) =\frac{h_4^{*}}{h_4}+\frac 2{h_4^{*}}\left[
h_4^{^{\prime \prime }}-\ddot h_4-\sin (h_4)\right] .
$$
The integral varieties (with energy--momentum sources and in vacuum cases)
are constructed by a corresponding redefinition of coefficients in formulas
from the previous paragraph.

\subsubsection{On some general properties of h--metrics depending on 2+1
variables}

By introducing a new variable $\beta =h_4^{*}/h_4$ the equation (\ref{heq})
transforms into
\begin{equation}
\label{heq1}\beta ^{*}+\frac 12\beta ^2-\frac{\beta h_3^{*}}{2h_3}-2\kappa
\Upsilon _1h_3=0
\end{equation}
which relates two functions $\beta \left( x^i,z\right) $ and $h_3\left(
x^i,z\right) .$ There are two possibilities: 1) to define $\beta $ (i. e. $%
h_4)$ when $\kappa \Upsilon _1$ and $h_3$ are prescribed and, inversely 2) to
find $h_3$ for given $\kappa \Upsilon _1$ and $h_4$ (i. e. $\beta );$ in
both cases one considers only ''*'' derivatives on $z$--variable with
coordinates $x^i$ treated as parameters.

\begin{enumerate}
\item  In the first case the explicit solutions of (\ref{heq1}) have to be
constructed by using the integral varieties of the general Riccati equation
\cite{kamke} which by a corresponding redefinition of variables, $%
z\rightarrow z\left( \varsigma \right) $ and $\beta \left( z\right)
\rightarrow \eta \left( \varsigma \right) $ (for simplicity, we omit
dependencies on $x^i)$ could be written in the canonical form
$$
\frac{\partial \eta }{\partial \varsigma }+\eta ^2+\Psi \left( \varsigma
\right) =0
$$
where $\Psi $ vanishes for vacuum gravitational fields. In vacuum cases the
Riccati equation reduces to a Bernoulli equation which (we can use the
former variables) for $s(z)=\beta ^{-1}$ transforms into a linear
differential (on $z)$ equation,
\begin{equation}
\label{heq1a}s^{*}+\frac{h_3^{*}}{2h_3}s-\frac 12=0.
\end{equation}

\item  In the second (inverse) case when $h_3$ is to be found for some
prescribed $\kappa \Upsilon _1$ and $\beta $ the equation (\ref{heq1}) is to
be treated as a Bernoulli type equation,
\begin{equation}
\label{heq2}h_3^{*}=-\frac{4\kappa \Upsilon _1}\beta (h_3)^2+\left( \frac{%
2\beta ^{*}}\beta +\beta \right) h_3
\end{equation}
which can be solved by standard methods. In the vacuum case the squared on $%
h_3$ term vanishes and we obtain a linear differential (on $z)$ equation.
\end{enumerate}

\subsubsection{A class of conformally equivalent h--metrics}

A particular interest presents those solutions of the equation (\ref{heq1})
which via 2D conformal transforms with a factor $\omega =\omega (x^i,z)$ are
equivalent to a diagonal h--metric on $x$--variables, i.e. one holds the
parametrization
\begin{equation}
\label{conf4d}h_3=\omega (x^i,z)\ a_3\left( x^i\right) \mbox{ and }%
h_4=\omega (x^i,z)\ a_4\left( x^i\right) ,
\end{equation}
where $a_3\left( x^i\right) $ and $a_4\left( x^i\right) $ are some arbitrary
functions (for instance, we can impose the condition that they describe some
2D soliton like or black hole solutions). In this case $\beta =\omega
^{*}/\omega $ and for $\gamma =\omega ^{-1}$ the equation (\ref{heq1})
trasforms into
\begin{equation}
\label{confeq}\gamma \ \gamma ^{**}=-2\kappa \Upsilon _1a_3\left( x^i\right)
\end{equation}
with the integral variety determined by
\begin{equation}
\label{confeqsol}z=\int \frac{d\gamma }{\sqrt{\left| -4k\Upsilon
_1a_3(x^i)\ln |\gamma |+C_1(x^i)\right| }}+C_2(x^i),
\end{equation}
where it is considered that the source $\Upsilon _1$ does not depend on $z.$

Finally, in this Section, we have shown that a large class of 4D solutions,
depending on two or three variables, of the Einstein equations can be
constructed as nonlinear superpositions of some 2D h--metrics defined by
locally anisotropic deformations of 2D sine--Gordon equations and of some
v--metrics generated in particular by solutions of Kadomtsev--Petviashvili
equations, or of (2+1) sine--Gordon, and associated Bernoulli type
equations. From a general viewpoint the v--metrics are defined by integral
varieties of corresponding Riccati and/or Bernoulli equations with respect
to $z$--variables with the h--coordinates $x^i$ treated as parameters.

\section{Effective Locally Anisotropic Soliton--Dilaton fields}

The formula for the h--component $\widehat{R}$ of scalar curvature (see (\ref
{dscalar}) and (\ref{ricci1})) of a h--metric (\ref{hansatz}), written for a
la--system, differs from the usual one for computation of curvature of 2D
metrics. That why additionally to the first term in (\ref{riccihs}) it is
induced the $\rho $--term (\ref{rho}). The aim of this Section is to prove
that the la--deformed singe--Gordon equation (\ref{riccihs}) could be
equivalently modelled by solutions of the usual 2D singe--Gordon equation
and an additional equation for a corresponding effective dilaton field (in
brief, by a soliton--dilaton field). We also analyze the 2D dilaton gravity
in connection with the sine--Gordon la--field theory.

\subsection{Generic locally anisotropic dilaton fields}

Let $\widetilde{g}_{ij}^\epsilon \left( x^i\right) $ be a 2D metric of
Lorentz (or Euclidean) signature for $\epsilon =-1$ (or $\epsilon = 1)$ with
a usual 2D scalar curvature $\widetilde{R}_{(\epsilon )}\left( x^i\right) .$
We also consider a conformally equivalent metric
\begin{equation}
\label{conf3}\underline{g}_{ij}^\epsilon \left( x^i\right) =\exp \omega
\left( x^i\right) \widetilde{g}_{ij}^\epsilon \left( x^i\right) .
\end{equation}
The scalar curvatures of 2D metrics from (\ref{conf3}) are related by the
formula
\begin{equation}
\label{curvund}e^\omega \underline{R}=\widetilde{R}_{(\epsilon )}+\triangle
_{(\epsilon )}\omega
\end{equation}
where $\triangle _{(\epsilon )}$ is the d'Alambert, $\epsilon =-1,$ (Laplace,%
$\epsilon =1$) operator.

In order to model a locally anisotropic 2D horizontal system via a locally
isotropic 2D gravity we consider that
$$
\widehat{R}=e^\omega \underline{R},\widetilde{R}_{(\epsilon )}=-2\widetilde{m%
}^2
$$
and
\begin{equation}
\label{poisson}\triangle _{(\epsilon )}\omega =\rho \left( x^i\right) .
\end{equation}
For a given 'tilded' metric, for instance, $\widetilde{g}_{ij}= diag(%
\widetilde{a},\widetilde{b})$ being a solution of 2D sine--Gordon equation (%
\ref{sg1}) with negative constant scalar curvature (see (\ref{curv1})), the
wave (Poisson) equation can be solved in explicit form by imposing
corresponding boundary conditions.

So, a 2D locally anisotropic h-space is equivalently modelled by a usual
curved 2D locally isotropic (pseudo) Riemannian space and effective
interactions with the generic locally anisotropic dilaton field $\Phi
_{(\omega )}=\exp \omega .$

\subsection{Locally anisotropic 2D dilaton gravity and sine--Gordon theory}

In the previous Subsection the conformal factor $\Phi _{(\omega )},$ in the
h--space, was introduced with the aim to compensate the local anisotropy,
induced from the v--space. The 2D h--gravity can be formulated
as a dilatonic
one related to a generalized, la--deformed, sine--Gordon model.

By using Weyl rescallings of h--metric (\ref{hansatz}) one can write the
general action for, in our case, the h--model (see the isotropic variant in
\cite{banks} and \cite{mann}),
\begin{equation}
\label{actdilaton}S^{[h]}\left[ g_{ij},\Phi \right] =\frac 1{2\pi }\int d^2x%
\sqrt{-g}[\Phi \widehat{R}+\varpi ^2V\left( \Phi \right) ],
\end{equation}
where the h--metric $g_{ij}$ has signature $\left( -1,1\right) ,$ $V\left(
\Phi \right) $ is an arbitrary function of the dilaton field $\Phi $ and $%
\varpi $ is the connection constant. The la--field equations derived from
this action are
\begin{eqnarray}
\widehat{R}=\widetilde{R}_{(-)}+\triangle _{(-)}\omega
 &=& -\varpi ^2\frac{dV}{d\Phi },\label{soldilsys} \\
D_iD_j\Phi -\frac{\varpi ^2}2g_{ij}V &=&0,  \nonumber
\end{eqnarray}
where $\widetilde{R}_{(-)}=2\widetilde{R}_1^1$ is defined by the
h--component of scalar curvature of type (\ref{riccihs}), when $\epsilon
=-1. $ In the locally isotropic limit this system of equations describes the
Cadoni theory \cite{cadoni}.

In consequence of the fact that the theory is invariant under coordinate
h--transforms $(x^1=t,x^2=x)$ we can introduce the h--metric
\begin{equation}
\label{trig2d}g^{[h]}=-\sin ^2\left( \frac v2\right) dt^2+\cos ^2\left(
\frac v2\right) dx^2,
\end{equation}
where $v=v(t,x)$ and rewrite the system (\ref{soldilsys}) as a system of
nonlinear partial differential equations in 2D Euclidean space,
\begin{eqnarray}
\ddot v+v^{\prime \prime } &= &
\left( -\rho +\frac{\varpi ^2}2\frac{dV}{d\Phi }\right) \sin v,
 \label{dilat1} \\
\ddot \Phi+\Phi ^{\prime \prime }&=&\frac{\varpi ^2}2V\cos v,
 \label{dilat2}
\end{eqnarray}
where the function $\rho $ is defined by the formula (\ref{rho}) for $%
\epsilon =-1,$ or, in equivalent form, by a generic la--dilaton given
 by (\ref{poisson}).

 The equation (\ref{dilat1}) reduces to the deformed sine--Gordon equation
(\ref{sgla}), for $\epsilon =-1,$ if $V=\Phi $, or for constant
 configurations $\Phi _0$ for which $V(\Phi _0)=0$ and \\
 $dV/d\Phi |_{\Phi _0}>0.$

For soliton--dilaton la--configurations it is more convenient to consider
the action
\begin{equation}
\label{actdilaton1}S=\frac 12\int d^2x\left[ \Phi \left( \triangle
_{(-)}v+\rho \right) -\frac{\varpi ^2}2V\sin v\right] ,
\end{equation}
given in the 2D Minkowski space, where $\triangle _{(-)}v=\ddot v+v^{\prime
\prime }$ and $\rho =\triangle _{(-)}\omega .$ Extremizing the action (\ref
{actdilaton1}) we obtain the field equations (\ref{dilat1}) and (\ref{dilat2}%
) as well from this action one follows the energy functional%
\begin{eqnarray}
E\left( v,\omega ,\Phi \right)& =& \frac 12\int\limits_{-\infty }^\infty dx[%
\dot \Phi\left( \dot v + \dot \omega\right) +\Phi ^{\prime }\left( v^{\prime
}+\omega ^{\prime }\right)  \nonumber \\
 &{}& +\frac{\varpi ^2}2V\sin v]. \nonumber
\end{eqnarray}

We note that instead of Lorentz type 2D h--metrics we can consider Euclidean
field equations by performing the Wick rotation $t\rightarrow it.$ We
conclude this Subsection by the remark that a complete correspondence
between locally anisotropic h--metrics and dilaton structures is possible if
additionally to trigonometric parametrizations of 2D metrics (\ref{trig2d})
one introduces hyperbolic parametizations
$$
g^{[h]}=-\sinh ^2\left( \frac v2\right) dt^2+\cosh ^2\left( \frac v2\right)
dx^2
$$
which results in sinh--Gordon models.

\subsection{Static locally anisotropic soliton--dilaton configurations}

Because the $\rho $--term (\ref{rho}) vanishes for constant values of
fields $v=2n\pi ,$ $n=0,\pm 1,\pm 2,...$ and $\Phi =\Phi _0$ the vacua of
the model (\ref{actdilaton}) is singled out like in the locally isotropic
case \cite{cadoni}, by conditions
\begin{equation}
\label{vacuum}V\left( \Phi _0\right) =0\mbox{ and }\frac{dV}{d\Phi }\left(
\Phi _0\right) >0.
\end{equation}
In order to focus on static deformations induced by soliton like solutions
we require $E\geq 0$ and
\begin{equation}
\label{limits}\lim \limits_{x\rightarrow \pm \infty .}v^{\prime }\rightarrow
0\mbox{ and }\lim \limits_{x\rightarrow \pm \infty .}\Phi ^{\prime
}\rightarrow 0.
\end{equation}

The static la--configurations of (\ref{dilat1}) and (\ref{dilat2}) are
giving by anholonomic deformation of isotropic ones and are given
by the system of equations%
\begin{eqnarray}
v^{\prime \prime }+\frac{\cos v-1}{\sin v}\left( v^{\prime }\right) ^2
 &=&\frac{\varpi ^2}2\sin v\ \frac{dV}{d\Phi },\label{statdef1} \\
\Phi ^{\prime \prime }=\frac{\varpi ^2}2V\left( \Phi \right) \cos v.%
\label{statdef2}
\end{eqnarray}
The first integrals of (\ref{statdef1}) and (\ref{statdef2}) are
$$
v^{\prime }=\varpi \frac{a_0}{\sqrt{2}}\sin ^2\frac v2\mbox{ and }\Phi
^{\prime }=\frac{\varpi \sqrt{2}}{a_0}\cot \frac v2
$$
which, after another integration, results in the solutions%
\begin{eqnarray}
\varpi \left( x-x_0\right) & = &
\pm \frac{a_0^2}{\sqrt{2}}\int d\Phi %
\sqrt{\frac{\Psi -c_0}{1-a_0^2\left( \Psi -c_0\right) }},%
\label{dilatunds} \\
\sin \frac v2 &=& \pm a_0\sqrt{\Psi -c_0}, \nonumber
\end{eqnarray}
where $\Psi =\Psi \left( \Phi \right) =\int_0^\Phi d\tau V\left( \tau
\right) ;$\ $a_0$ and $c_0=\Psi [\underline{\Phi }\left( \pm \infty \right)
] $ are integration constants.
 We emphasize that the formula (\ref{dilatunds}),
 following from a la--model, differs from that
 obtained in the Cadoni's locally isotropic theory \cite{cadoni}.

There are two additional two parameter solutions of (\ref{statdef1}) and (%
\ref{statdef1}) which are not contained in (\ref{dilatunds}). The first type
of solutions are those for constant $v$ field when
$$
v=n\pi \mbox{ and }\varpi \left( x-x_0\right) =\pm \int d\Phi \left[
(-1)^n\Psi -b_0\right] ^{-1/2},
$$
where $b_0=const.$ The second type of solutions are for constant dilaton
fields $\Phi _0$ and exists if there is at least one zero $\Phi =\Phi _0$
for $V\left( \Phi \right) .$ For $dV/d\Phi \left( \Phi _0\right) >0$ the
equations reduce to the usual sine--Gordon equations
$$
v^{\prime \prime }=\frac{\varpi ^2}2\frac{dV}{d\Phi }\mid _{\Phi _0}\sin v.
$$

Note that the model (\ref{actdilaton1}) admits static soliton solutions,
approaching for $x\rightarrow \pm \infty $ the constant field configuration $%
v=2\pi n;\ n=\pm 1,$ with $\Phi _0=\Phi \left( \pm \infty \right) ,$ $%
V\left( \Phi _0\right) =0$ and $dV/d\Phi |_{\Phi _0}>0.$

\subsection{Topology of locally anisotropic soliton--dilatons}

If we suppose that la--deformations do not change the spacetime topology,
the conditions (\ref{limits}) imply that every soliton solution tends
asymptotically to one of vacuum configurations (\ref{vacuum}) which could
be considered for both locally anisotropic and isotropic systems (see \cite
{cadoni}). The admissible number of solitons to be la--deformed without
changing of topology is determined by the number of ways in which the
points $x=\pm \infty $ (the zero sphere) can be mapped into the manifold of
the mentioned constant--field configurations (\ref{vacuum}) characterized by
the homotopy group
$$
\pi _0\left( \frac{Z\times Z_2}{Z_2}\right) =\pi _0\left( Z\right) =Z,
$$
when $G=Z\times Z_2$ is the invariance group for a generic $V,$ and $Z$ and $%
Z_2$ are respectively the infinite discrete group translations and the
finite group of inversions of the field $v,$ parametrized by $v\rightarrow
v+2\pi n,$ $v\rightarrow -v.$ This result holds for usual sine--Gordon
systems, as well by la--generalizations given by the action (\ref
{actdilaton1}) and la--field equations (\ref{dilat1}) and (\ref{dilat2}).

For soliton like theories it is possible the definition of conserved
currents
$$
J_{(v)}^i=\epsilon ^{ij}\delta _iv\mbox{ and }J_{(\Phi )}^i=\epsilon
^{ij}\delta _i\Phi ,
$$
where $\epsilon ^{ij}=-\epsilon ^{ji}$ and the 'elongated' (in la--case)
partial derivatives $\delta _i$ are given by (\ref{dder}). The associated
topological charges on a fixed la--background are
\begin{eqnarray}
Q_{(v)} &=&
\frac 1{2\pi }\int\limits_{-\infty }^\infty dxJ_{(v)}^1=\frac 1{2\pi
}\left[ v\left( \infty \right) -v\left( -\infty \right) \right] , \nonumber
\\
Q_{(\Phi )} &=&
\frac 1{2\pi }\int\limits_{-\infty }^\infty dxJ_{(\Phi )}^1=%
\frac 1{2\pi }\left[ \Phi \left( \infty \right) -\Phi \left( -\infty \right)
\right] . \nonumber
\end{eqnarray}

The topological properties of la--backgroudns are characterized by the
topological current and charge of la--dilaton $\Phi _{(\omega )}=\exp \omega $
defined by a solution of Poisson equation (\ref{poisson}). The
corresponding formulas are
\begin{eqnarray}
J_{(e^\omega )}^i &=&\epsilon ^{ij}\delta _ie^\omega ,\label{topsourse2}
\\
Q_{(e^\omega )}&=&
\frac 1{2\pi }\int\limits_{-\infty }^\infty dxJ_{(e^\omega
)}^1=\frac 1{2\pi }\left[ \exp \omega \left( \infty \right) -\exp \omega
\left( -\infty \right) \right] .  \nonumber
\end{eqnarray}
If $J_{(e^\omega )}^i$ and $Q_{(e^\omega )}$ are non-trivial we can conclude
that our soliton--dilaton system was topologically changed under
la--deformations.

\section{Locally Anisotropic Black Holes and Solitons}

In this Section we analyze the connection between h--metrics
 describing effective 2D black la--hole solutions (with
 parameters defined by v--components of a diagonal 4D energy--momentum
 d--tensor) and 2D soliton la--solutions obtained in the previous two
 Sections.

\subsection{A Class 1 black hole solutions}

Let us consider a static h--metric of type (\ref{dm2drmh}), for which $%
g_1=-\alpha \left( r\right) $ and $g_2=1/\alpha \left( r\right) $ for a
function on necessary smooth class $\alpha $ on h--coordinates $x^1=T$
and $x^2=r.$ Putting these values of h--metric into (\ref{ricci1}) we
compute
$$
R_1^1=R_2^2=-\frac 12\alpha ^{\prime \prime }.
$$
Considering the 2D h--subspace to be of constant negative scalar curvature,%
$$
\widehat{R}=2R_1^1=-\widetilde{m}^2,
$$
and that the Einstein la--equations (\ref{einsteq3a}) are satisfied we
obtain the relation
\begin{equation}
\label{bheq1}\alpha ^{\prime \prime }=\widetilde{m}^2=\kappa \Upsilon
_3^3=\kappa \Upsilon _4^4,
\end{equation}
which for a diagonal energy--momentum d--tensor with $\kappa \Upsilon
_3^3=kp_3+\lambda /2$ transforms into%
$$
\widetilde{m}^2=kp_3+\frac \lambda 2.
$$

The solution of (\ref{bheq1}) is written in the form $\alpha =\left(
\widetilde{m}^2r^2-M\right) $ which defines a 2D h--metric
\begin{equation}
\label{bh1}ds_{(h)}^2=-\left( \widetilde{m}^2r^2-M\right) dT^2+\left(
\widetilde{m}^2r^2-M\right) ^{-1}dr^2
\end{equation}
being similar to a black hole solution in 2D Jackiw--Teitelboim gravity
\cite{jackiw} and display many of attributes of black holes \cite
{mann,gegenberg1,lemos1} with that difference that the constant $\widetilde{m%
}$ is defined by 4D physical values in v--subspace and for definiteness of
the theory the h--metric should be supplied with a v--component.

The parameter $M,$ the mass observable, is the analogue of the
Arnowitt--Deser--Misner (ADM) mass in general relativity. If we associate the
h--metric (\ref{bh1}) to a 2D model of Jackiw--Teitelboim la--gravity, given
by the action
$$
I_{JT}[\phi ,g]=\frac 1{2G_{(2)}}\int_Hd^2x\sqrt{\left| g\right| }\phi
\left( \widehat{R}+2\widetilde{m}^2\right)
$$
where $\widehat{R}$ is the h--component of the Ricci d--curvature, $\phi $
is the dilaton field and $G_{(2)}$ is the gravitational coupling constant in
2D, we should add to (\ref{bheq1}) the field equation for $\phi ,$%
$$
\left( D_iD_j-\widetilde{m}^2g_{ij}\right) \phi =0
$$
which has the solution
$$
\phi =c_1\widetilde{m}r
$$
with coupling constant $c_1,$ (we can consider $c_1=1$ for the vacuum
configurations $\phi =\widetilde{m}r$ as $r\rightarrow \infty ).$
In this case
the mass observable is connected with the dilaton as
\begin{equation}
\label{mass1}M=-\widetilde{m}^{-2}\left| D\phi \right| ^2+\phi ^2.
\end{equation}

Clearly this model is with local anisotropy because the value $\widehat{R}$
is defined in a la--manner and not as a usual scalar curvature in 2D gravity.

\subsection{La--deformed soliton--dilaton systems and black la--holes}

Suppose we have a h--metric (\ref{trig2d}) which must solve the equations%
\begin{eqnarray}
\ddot v+v^{\prime \prime } &= &
\left( -\rho +{\tilde m}^2 \right) \sin v,
 \label{dilat1a} \\
\ddot \phi+\phi ^{\prime \prime }&=& {\tilde m}^2\cos v.
 \label{dilat2a}
\end{eqnarray}
These equations are a particular case of the system (\ref{dilat1}) and (\ref
{dilat2}). For $\rho =0$ such equations were investigated in
\cite{gegenberg}.
In the previous Section we concluded that every 2D la--system can be
equivalently modelled in an isotropic space by considering an effective
interaction with la--dilaton field. The same considerations hold good
for 2D la--spaces with that remark that the dilaton field $\phi $ must be
composed from a component satisfying the equation (\ref{dilat2a}) and
another component defined from the Poisson equation (\ref{poisson}).

Having a dilaton field $\phi \left( t,x\right) $ we can introduce a new
''radial coordinate''%
$$
r\left( t,x\right) :=\phi /\widetilde{m}
$$
which (being substituted into h--metric (\ref{trig2d})) results in the
horizontal 2D metric
\begin{equation}
\label{bh2}ds_{[h]}^2=-\widetilde{m}^{-2}\left| D\phi \right| ^2dT^2+%
\widetilde{m}^2\left| D\phi \right| ^{-2} dr^2,
\end{equation}
where
$$
dT\doteq \left| D\phi \right| ^{-2}\left( \phi ^{\prime }\tan \frac v2\ dt+%
\dot \phi \cot \frac v2\ dx\right) .
$$
The metric (\ref{bh2}) is the same as (\ref{bh1}) because the mass
observable is defined by (\ref{mass1}).

\subsection{The geometry of black la--holes and deformed one--soliton
solutions}

The one soliton solution of the Euclidean sine--Gordon equation can be
written as
\begin{equation}
\label{soliton2}v(t,x)=4\tan ^{-1}\left\{ \exp \left[ \pm \widetilde{m}%
\gamma \left( x-st-x_{(0)}\right) \right] \right\} ,
\end{equation}
where $\gamma =1/\sqrt{1+s^2},$ $s$ is the spectral parameter and the
integration constant $x_{(0)}$ is considered, for simplicity, zero. The
solution with signs $+/-$ gives a soliton/anti--soliton configuration.
  Let us demonstrate that a corresponding black la--hole can be constructed.
 Putting (\ref{soliton2}) into the h--metric (\ref{trig2d}) we obtain
  a Lorentzian one--soliton 2D metric
$$
ds_{[1-sol]}^2=-\sec h^2\xi \ dt^2+\tanh ^2\xi \ dx^2
$$
where
$$
\xi \doteq \widetilde{m}\gamma \left( x-st\right) .
$$
In a similar fashion we can compute (by using the function (\ref{soliton2}))
the la--deformation $\rho $ (\ref{rho}) and effective la--dilaton $\Phi
_\omega =\exp \omega $ which follow from the Poisson equation (\ref{poisson}%
). In both cases of dilaton equations we are dealing with linear partial
differential equations. A combination of type
\begin{equation}
\label{dilaton2}\phi =\phi _{[0]}\dot v+\phi _{[1]}v^{\prime }
\end{equation}
for arbitrary constants $\phi _{[0]}$ and $\phi _{[1]}$ satisfies the
linearized sine--Gordon equation and because for the function (\ref{soliton2}%
)%
$$
\dot v=\mp 4\widetilde{m}\gamma s\ \sec h\ \xi =-sv^{\prime }
$$
we can put in (\ref{dilaton2}) $\phi _{[1]}=0$ and following a Hamiltonian
analysis (in order to have compatibility with the locally isotropic case
\cite{gegenberg}; for a black hole mass $M=s^2$ with corresponding ADM\
energy $E=\widetilde{m}^2s^2/\left( 2G_{(2)}\right) )$ we set $\phi
_{[0]}=1/\left( 4\widetilde{m}\gamma ^2\right) $ so that
$$
\phi =\left| \frac s\gamma \right| \sec h\ \xi
$$
is chosen to make $\phi $ positive. In consequence, the black hole
coordinates $\left( r,T\right) $ (la--deformations reduces to
reparametrization of such coordinates) are defined by
$$
r=\phi /\widetilde{m}=\frac s{\widetilde{m}\gamma }\sec h\ \xi
$$
and
$$
dT=\left| s\widetilde{m}\right| ^{-1}\left[ dt-\frac{s\tanh ^2\ \xi }{%
\widetilde{m}\gamma \left( \sec h^2\ \xi -s^2\tanh ^2\ \xi \right) }d\xi
\right] .
$$
 With respect to
 these coordinates the obtained black hole metric is of the form%
$$
ds_{[bh]}^2=-\left( \widetilde{m}^2r^2-\widetilde{m}^2s^4\right) dT^2+\left(
\widetilde{m}^2r^2-\widetilde{m}^2s^4\right) ^{-1}dr^2
$$
which describes a Jackiw--Teitelboim black hole with mass parameter (\ref
{mass1})
$$
M_{1sol}=\widetilde{m}^2s^4,
$$
defined by the corresponding component $\Upsilon _3^3=\Upsilon _4^4$ of
energy--momentum d--tensor in v--space and spectral parameter $s$ of the one
soliton background, and event horizon at $\phi =\phi _H=s^2.$

In a similar fashion we can use instead of the function (\ref{soliton2}) a
two and, even multi--, soliton background. The calculus is similar to the
locally isotropic case \cite{gegenberg}, having some redefinitions of black
hole coordinates if it is considered that la--deformations do not change the
h--spaces topology, i.e the la--gravitational topological source and charge (%
\ref{topsourse2}) vanishes.

\section{3D Black La--Holes}

Let us analyze some basic properties of 3D spacetime $V^{(2+1)}$ provided
with d--metrics of type
\begin{equation}
\label{dmetr3}\delta s^2=g_1\left( x^k\right) \left( dx^1\right)
^2+g_2\left( x^k\right) \left( dx^2\right) ^2+h_3(x^i,z)\left( \delta
z\right) ^2,
\end{equation}
where $x^k$ are 2D coordinates, $y^3=z$ is the anisotropic coordinate and
$$
\delta z=dz+N_i^3(x^k,z)dx^i.
$$
The N--connection coefficients are given by some functions on $x^i$ and $z,$%
\begin{equation}
\label{ncoef3}N_1^3=q_1(x^i,z),\ N_2^3=q_2(x^i,z).
\end{equation}

The non--trivial components of the Ricci d--tensor (\ref{dricci}) are%
\begin{eqnarray}
&R_1^1&=R_2^2=\frac 1{2g_1g_2} \ [-(g_1^{^{\prime \prime }}
+{\ddot g}_2)  \label{ricci1_3} \\
&{}& +\frac 1{2g_2}\left( {\dot g}%
_2^2+g_1^{\prime }g_2^{\prime }\right) +\frac 1{2g_1}\left( g_1^{\prime \ 2}+%
\dot g_1\dot g_2\right) ]; \nonumber
\end{eqnarray}
\begin{equation}
\label{ricci3_3}P_{3i} = \frac{q_i}2[\left( \frac{h_3^{*}}{h_3}\right) ^2-
\frac{h_3^{**}}{h_3} ]
\end{equation}
(for 3D the component $S_3^3\equiv 0,$ see (\ref{ricci2})).

The curvature scalar $\overleftarrow{R}$ (\ref{dscalar}) is $\overleftarrow{R%
}=\widehat{R}=2R_1^1.$

The system of Einstein equations (\ref{einsteq2}) transforms into
\begin{eqnarray}
R_1^1&=&-\kappa \Upsilon _3^3,
\label{einsteq3a3} \\
P_{3i}&=& \kappa \Upsilon _{3i}, \label{einsteq3c3}
\end{eqnarray}
which is compatible for if the 3D matter is in a state when for the
energy--momentum d--tensor $\Upsilon _\beta ^\alpha $ one holds $\Upsilon
_1^1=\Upsilon _2^2=0$ and the values of $R_1^1,P_{3i},$ are taken
respectively from (\ref{ricci1_3}) and (\ref{ricci3_3}).

By using the equation (\ref{einsteq3c3}) we can define the N--coefficients (%
\ref{ncoef3}), $q_i(x^k,z),$ if the function $h_3(x^k,z)$ and the components
$\Upsilon _{3i}$ of the energy--momentum d--tensor are given. We note that
the equations (\ref{ricci3_3}) are solved for arbitrary functions $%
h_3=h_3(x^k)$ and $q_i=q_i(x^k,z)$ if $\Upsilon _{3i}=0$ and in this case
the component of d--metric $h_3(x^k)$ is not contained in the system of 3D
field equations.

\subsection{Static elliptic horizons}

Let us consider a class of 3D d-metrics which local anisotropy which are
similar to Banados--Teitelboim--Zanelli (BTZ) black holes \cite{btz}.

The d--metric is parametrized
\begin{equation}
\label{dim3}\delta s^2=g_1\left( \chi ^1,\chi ^2\right) (d\chi ^1)^2+\left(
d\chi ^2\right) ^2-h_3\left( \chi ^1,\chi ^2,t\right) \ \left( \delta
t\right) ^2,
\end{equation}
where $\chi ^1=r/r_h$ for $r_h=const,$ $\chi ^2=\theta /r_a$ if $r_a=\sqrt{%
|\kappa \Upsilon _3^3|}\neq 0$ and $\chi ^2=\theta $ if $\Upsilon _3^3=0,$ $%
y^3=z=t,$ where $t$ is the time like coordinate. The Einstein equations (\ref
{einsteq3a3}) and (\ref{einsteq3c3}) transforms respectively into
\begin{equation}
\label{hbh1a3}\frac{\partial ^2g_1}{\partial (\chi ^2)^2}-\frac
1{2g_1}\left( \frac{\partial g_1}{\partial \chi ^2}\right) ^2-2\kappa
\Upsilon _3^3g_1=0
\end{equation}
and
\begin{equation}
\label{hbh1c3}\left[ \frac 1{h_3}\frac{\partial ^2h_3}{\partial z^2}-\left(
\frac 1{h_3}\frac{\partial h_3}{\partial z}\right) ^2\right] q_i=-\kappa
\Upsilon _{3i}.
\end{equation}
By introducing new variables
\begin{equation}
\label{p-var3}p=g_1^{\prime }/g_1\mbox{ and }s=h_3^{*}/h_3
\end{equation}
where the 'prime' in this subsection denotes the partial derivative $%
\partial /\chi ^2,$ the equations (\ref{hbh1a3}) and (\ref{hbh1c3})
transform into
\begin{equation}
\label{hbh2a3}p^{\prime }+\frac{p^2}2+2\epsilon =0
\end{equation}
and
\begin{equation}
\label{hbh2c3}s^{*}q_i=\kappa \Upsilon _{3i},
\end{equation}
where the vacuum case should be parametrized for $\epsilon =0$ with $\chi
^i=x^i$ and $\epsilon =1(-1)$ for the signature $1(-1)$ of the anisotropic
coordinate.

A class of solutions of 3D Einstein equations for arbitrary $q_i=q_i(\chi
^k,t)$ and $\Upsilon _{3i}=0$ is obtained if $s=s(\chi ^i).$ After
integration of the second equation from (\ref{p-var3}), we find
\begin{equation}
\label{hbh2c3s}h_3(\chi ^k,t)=h_{3(0)}(\chi ^k)\exp \left[ s_{(0)}\left(
\chi ^k\right) t\right]
\end{equation}
as a general solution of the system (\ref{hbh2c3}) with vanishing right
part. Static solutions are stipulated by $q_i=q_i(\chi ^k)$ and $%
s_{(0)}(\chi ^k)=0.$

The integral curve of (\ref{hbh2a3}), intersecting a point $\left( \chi
_{(0)}^2,p_{(0)}\right) ,$ considered as a differential equation on $\chi ^2$
is defined by the functions \cite{kamke}%
\begin{eqnarray}
p &=&
\frac{p_{(0)}}{1+\frac{p_{(0)}}2 %
\left( \chi ^2-\chi _{(0)}^2\right) },\qquad   \epsilon =0; \label{eq3a} \\%
p & = & \frac{p_{(0)}-2\tanh \left( \chi ^2- %
\chi _{(0)}^2\right) }{1+\frac{p_{(0)}}2  %
  \tanh \left( \chi ^2-\chi _{(0)}^2\right) },\qquad  %
  \epsilon >0; \label{eq3b}  \\    %
 p & = & \frac{p_{(0)}-2\tan \left( \chi ^2-\chi _{(0)}^2\right) }%
 {1+\frac{p_{(0)}}2\tan \left( \chi ^2-\chi _{(0)}^2\right) },\qquad %
 \epsilon <0.   \label{eq3c} %
\end{eqnarray}

Because the function $p$ depends also parametrically on variable $\chi ^1$
we must consider functions $\chi _{(0)}^2=\chi _{(0)}^2\left( \chi ^1\right)
$ and $p_{(0)}=p_{(0)}\left( \chi ^1\right) .$

For simplicity, here we elucidate the case $\epsilon <0.$ The general
formula for the non--trivial component of h--metric is to be obtained after
integration on $\chi ^1$ of (\ref{eq3c}) (see formula (\ref{p-var3}))%
\begin{eqnarray}
& g_1\left( \chi ^1,\chi ^2\right) &= g_{1(0)} \left( \chi ^1\right) 
\times \nonumber \\
 &{} & \left\{ \sin [\chi ^2-\chi _{(0)}^2
\left( \chi ^1\right) ]+\arctan \frac 2{p_{(0)}\left(
\chi ^1\right) }\right\} ^2, \nonumber
\end{eqnarray}
for $p_{(0)}\left( \chi ^1\right) \neq 0,$ and
\begin{equation}
\label{btzlh3}g_1\left( \chi ^1,\chi ^2 \right) =g_{1(0)}\left( \chi
^1\right) \ \cos ^2[\chi ^2-\chi _{(0)}^2\left( \chi ^1\right) ]
\end{equation}
for $p_{(0)}(\chi ^1) =0,$ where $g_{1(0)}(\chi^1),
\chi _{(0)}^2(\chi ^1) $ and $p_{(0)}(\chi^1) $
are some functions of necessary smoothness class on variable $%
\chi ^1=x^1/\sqrt{\kappa \varepsilon },$ when $\varepsilon $ is the energy
density. If we consider $\Upsilon _{3i}=0$ and a non--trivial diagonal
components of energy--momentum d--tensor, $\Upsilon _\beta ^\alpha
=diag[0,0,-\varepsilon],$ the N--connection coefficients $q_i(\chi ^i, t)$
could be arbitrary functions.

For simplicity, in our further considerations we shall apply the solution (%
\ref{btzlh3}).

The d--metric (\ref{dim3}) with the coefficients (\ref{btzlh3}) and (\ref
{hbh2c3s}) gives a general description of a class of solutions with generic
local anisotropy of the Einstein equations (\ref{einsteq2}).

Let us construct static black la--hole solutions for $s_{(0)}\left( \chi
^k\right) =0$ in (\ref{hbh2c3s}).

In order to construct an explicit la--solution we choose some
coefficients $h_{3(0)}(\chi ^k),g_{1(0)}\left( \chi ^1\right) $ and $\chi
_0\left( \chi ^1\right) $ following some physical considerations. For instance,
the Schwarzschild solution is selected from a general 4D metric with some
general coefficients of static, spherical symmetry by relating the radial
component of metric with the Newton gravitational potential. In this
section, we construct a locally anisotropic BTZ like solution by supposing
that it is conformally equivalent to the BTZ solution if one neglects
anisotropies on angle $\theta ),$
$$
g_{1(0)}\left( \chi ^1\right) =\left[ r\left( -M_0+\frac{r^2}{l^2}\right)
\right] ^{-2},
$$
where $M_0=const>0$ and $-1/l^2$ is a constant (which is to be considered
the cosmological from the locally isotropic limit. The time--time
coefficient of d--metric is chosen
\begin{equation}
\label{btzlva3}h_3\left( \chi ^1,\chi ^2\right) =r^{-2}\lambda _3\left( \chi
^1,\chi ^2\right) \cos ^2[\chi ^2-\chi _{(0)}^2\left( \chi ^1\right) ].
\end{equation}

If we chose in (\ref{btzlva3})
$$
\lambda _3={(-M_0+\frac{r^2}{l^2})}^2,
$$
when the constant
$$
r_h=\sqrt{M_0}l
$$
defines the radius of a circular horizon, the la--solution is conformally
equivalent, with the factor $r^{-2}\cos ^2[\chi ^2-\chi _{(0)}^2\left( \chi
^1\right) ], $ to the BTZ solution embedded into a anholonomic background
given by arbitrary functions $q_i(\chi ^i,t)$ which are defined by some
initial conditions of gravitational la--background polarization.

A more general class of la--solutions could be generated if we put, for
instance,
$$
\lambda _3\left( \chi ^1,\chi ^2\right) ={(-M}\left( \theta \right) {+\frac{%
r^2}{l^2})}^2,
$$
with
$$
{M}\left( \theta \right) =\frac{M_0}{(1+e\cos \theta )^2},
$$
where $e<1.$ This solution has a horizon, $\lambda _3=0,$ parametrized by an
ellipse
$$
r=\frac{r_h}{1+e\cos \theta }
$$
with parameter $r_h$ and eccentricity $e.$

We note that our solution with elliptic horizon was constructed for a
diagonal energy--momentum d-tensor with non--trivial energy density but
without cosmological constant. On the other hand the BTZ solution was
constructed for a generic 3D cosmological constant. There is not a
contradiction here because the la--solutions can be considered for a
d--tensor $\Upsilon _\beta ^\alpha =diag[p_1-1/l^2,p_2-1/l^2,-\varepsilon
-1/l^2]$ with $p_{1,2}=1/l^2$ and $\varepsilon _{(eff)}=\varepsilon +1/l^2$
(for $\varepsilon =const$ the last expression defines the effective constant
$r_a).$ The locally isotropic limit to the BTZ black hole could be realized
after multiplication on $r^2$ and by approximations $e\simeq 0,$ $\cos
[\theta -\theta _0\left( \chi ^1\right) ]\simeq 1$ and $q_i(x^k,t)\simeq 0.$

\subsection{Oscillating elliptic horizons}

The simplest way to construct 3D solutions of the Einstein equations with
oscillating in time horizon is to consider matter states with constant
nonvanishing values of $\Upsilon _{31}=const.$ In this case the coefficient $%
h_3$ could depend on $t$--variable. For instance, we can chose such initial
values when
\begin{equation}
\label{btzlva3osc1}h_3(\chi ^1,\theta ,t)=r^{-2}\left( -M\left( t\right) +%
\frac{r^2}{l^2}\right) \cos ^2[\theta -\theta _0\left( \chi ^1\right) ]
\end{equation}
with
$$
M=M_0\exp \left( -\widetilde{p}t\right) \sin \widetilde{\omega }t,
$$
or, for an another type of anisotropy,
\begin{eqnarray}
\label{btzlva3osc2}h_3(\chi ^1,\theta ,t)&=&
r^{-2}\left( -M_0+\frac{r^2}{l^2}\right) \times \nonumber \\
&{}& \cos ^2\theta \ \sin ^2[\theta -\theta _0\left( \chi ^1,t\right) ]
 \nonumber
\end{eqnarray}
with
$$
\cos \theta _0\left( \chi ^1,t\right) =e^{-1}\left( \frac{r_a}r\cos \omega
_1t-1\right) ,
$$
when the horizon is given parametrically, %
$$
r=\frac{r_a}{1+e\cos \theta }\cos \omega _1t,
$$
where the new constants (comparing with those from the previous subsection)
are fixed by some initial and boundary conditions as to be $\widetilde{p}>0,$
and $\widetilde{\omega }$ and $\omega _1$ are treated as some real numbers.

For a prescribed value of $h_3(\chi ^1,\theta ,t)$ with non--zero source $%
\Upsilon _{31},$ in the equation (\ref{einsteq3c3}), we obtain
\begin{equation}
\label{ncon3osc}q_1(\chi ^1,\theta ,t)=\kappa \Upsilon _{31}\left( \frac{%
\partial ^2}{\partial t^2}\ln |h_3(\chi ^1,\theta ,t)|\right) ^{-1}.
\end{equation}

A solution (\ref{dmetr3}) of the Einstein equations (\ref{einsteq3a3}) and (%
\ref{einsteq3c3}) with $g_2(\chi ^i)=1$ and $g_1(\chi ^1,\theta )$ and $%
h_3(\chi ^1,\theta ,t)$ given respectively by formulas (\ref{btzlh3}) and (%
\ref{btzlva3osc1}) describe a 3D evaporating black la--hole solution with
circular oscillating in time horizon. An another type of solution, with
elliptic oscillating in time horizon, could be obtained if we choose (\ref
{btzlva3osc2}). The non--trivial coefficient of the N--connection must be
computed following the formula (\ref{ncon3osc}).

\section{4D Locally Anisotropic Black Holes}

\subsection{Basic properties}

The purpose of this Section is the
 construction of d--metrics of Class 2, or 3 (see
(\ref{dm2drmh1}) and (\ref{dm2drmv1}), or (\ref{dm2drmh2}) and (\ref
{dm2drmv2})) which are conformally equivalent to some la--deformations of
black hole, disk, torus
 and cylinder like solutions. We shall analyze 4D d-metrics
of type
\begin{eqnarray}
\label{dmetr4}\delta s^2 &= & g_1\left( x^k\right) \left( dx^1\right) ^2+
\left(dx^2\right) ^2  \\
&{}& + h_3(x^i,z)\left( \delta z\right) ^2 +
h_4(x^i,z)\left( \delta y^4 \right) ^2.
 \nonumber
\end{eqnarray}

The Einstein equations (\ref{einsteq3a}) with the Ricci h--tensor (\ref
{ricci1}) and energy momentum d--tensor (\ref{dem1}), or (\ref{dem2}),
transforms into
\begin{equation}
\label{hbh1}\frac{\partial ^2g_1}{\partial (x^1)^2}-\frac 1{2g_1}\left(
\frac{\partial g_1}{\partial x^1}\right) ^2-2\kappa \Upsilon _3^3g_1=0.
\end{equation}
By introducing the coordinates $\chi ^i=x^i/\sqrt{|\kappa
\Upsilon _3^3|}$ for the Class 3 (2) d--metrics and the variable $%
p=g_1^{\prime }/g_1,$ where by 'prime' in this Section is considered the
partial derivative $\partial /\chi ^2,$ the equation (\ref{hbh1}) transforms
into
\begin{equation}
\label{hbh2}p^{\prime }+\frac{p^2}2+2\epsilon =0,
\end{equation}
where the vacuum case should be parametrized for $\epsilon =0$ with $\chi
^i=x^i$ and $\epsilon =1(-1)$ for Class 2 (3) d--metrics. The equations (\ref
{hbh1}) and (\ref{hbh2}) are, correspondingly, equivalent to the equations (%
\ref{hbh1a3}) and (\ref{hbh2a3}) with that difference that in this Section
we are dealing with 4D coefficients and values. The solutions for the
h--metric are parametrized like (\ref{eq3a}), (\ref{eq3b}), and (\ref{eq3c})
and the coefficient $g_1(\chi ^i)$ is given by a similar to
(\ref{btzlh3})
formula (for simplicity, here we elucidate the case $\epsilon <0)$ which for
$p_{(0)}\left( \chi ^1\right) =0$ transforms into
\begin{equation}
\label{btzlh4}g_1\left( \chi ^1,\chi ^2\right) =g_{1(0)}\left( \chi
^1\right) \ \cos ^{2}[\chi ^2-\chi _{(0)}^2\left( \chi ^1\right) ],
\end{equation}
where $g_1\left( \chi ^1\right) ,\chi _{(0)}^2\left( \chi ^1\right) $ and $%
p_{(0)}\left( \chi ^1\right) $ are some functions of necessary smoothness
class on variable $\chi ^1=x^1/\sqrt{\kappa \varepsilon },$ $\varepsilon $
is the energy density. The coefficients $g_1\left( \chi ^1,\chi ^2\right) $ (%
\ref{btzlh4}) and $g_2\left( \chi ^1,\chi ^2\right) =1$ define a h--metric
of Class 3 (\ref{dm2drmh2}) with energy--momentum d--tensor (\ref{dem2}).
The next step is the construction of h--components of d--metrics for
different classes of symmetries of anisotropies.

Now, let us consider the system of equations (\ref{einsteq3b}) with the
vertical Ricci d--tensor component (\ref{ricci2}) which are satisfied by
arbitrary functions
\begin{equation}
\label{hdm2var}h_3=a_3(\chi ^i)\mbox{ and }h_4=a_4(\chi ^i).
\end{equation}
If v--metrics depending on three coordinates are introduced, $h_a=h_a(\chi
^i,z),$ the v--components of the Einstein equations transforms into (\ref
{heq}) which reduces to (\ref{heq1}) for prescribed values of $h_3(\chi
^i,z),\,$ and, inversely, to (\ref{heq2}) if $h_4(\chi ^i,z)$ is prescribed.
For h--metrics being conformally equivalent to (\ref{hdm2var}) (see
transforms (\ref{conf4d})) we are dealing to equations of type (\ref{confeq}%
) with integral varieties (\ref{confeqsol}).

\subsection{Rotation Hypersurfaces Horizons}

We proof that there are static black hole and cylindrical like solutions of
the Einstein equations with horizons being some 3D rotation hypersurfaces.
The space components of corresponding d--metrics are conformally equivalent
to some locally anisotropic deformations of the spherical symmetric
Schwarzschild and cylindrical Weyl solutions. We note that for some classes
of solutions the local anisotropy is contained in non--perturbative
anholonomic structures.

\subsubsection{Rotation ellipsoid configuration}

There are two types of rotation ellipsoids, elongated and flattened ones. We
examine both cases  of such horizon configurations

\vskip0.2cm

\paragraph{Elongated rotation ellipsoid coordinates:}

${~}$\\ ${\qquad}$
 An elongated rotation ellipsoid hypersurface is given by the formula \cite
{korn}
\begin{equation}
\label{relhor}\frac{\widetilde{x}^2+\widetilde{y}^2}{\sigma ^2-1}+\frac{%
\widetilde{z}^2}{\sigma ^2}=\widetilde{\rho }^2,
\end{equation}
where $\sigma \geq 1$ and $\widetilde{\rho }$ is similar to the radial
coordinate in the spherical symmetric case.

The space 3D coordinate system is defined%
\begin{eqnarray}
\widetilde{x} &=&\widetilde{\rho}\sinh u\sin v\cos \varphi ,\
\widetilde{y}=\widetilde{\rho}\sinh u\sin v\sin \varphi ,\nonumber \\
\widetilde{z}&=& \widetilde{\rho}\ \cosh u\cos v, \nonumber
\end{eqnarray}
where $\sigma =\cosh u,(0\leq u<\infty ,\ 0\leq v\leq \pi ,\ 0\leq \varphi
<2\pi ). $\ The hypersurface metric is
\begin{eqnarray}
g_{uu} &=& g_{vv}=\widetilde{\rho}^2\left( \sinh ^2u+\sin ^2v\right) ,
 \label{hsuf1} \\
g_{\varphi \varphi } &=&\widetilde{\rho}^2\sinh ^2u\sin ^2v.
 \nonumber
\end{eqnarray}

Let us introduce a d--metric
\begin{eqnarray}
\label{rel1}\delta s^2 &=& g_1(u,v)du^2+dv^2 + \\ &{}&
 h_3\left( u,v,\varphi \right) \left( \delta t\right) ^2+
h_4\left( u,v,\varphi \right) \left( \delta\varphi \right) ^2,
 \nonumber
\end{eqnarray}
where $\delta t$ and $\delta \varphi $ are N--elongated differentials.

As a particular solution (\ref{btzlh4}) for the h--metric we choose the
coefficient
\begin{equation}
\label{relh1h}g_1(u,v)=\cos ^2v.
\end{equation}
The $h_3(u,v,\varphi )=h_3(u,v,\widetilde{\rho }
\left( u,v,\varphi \right) )$
is considered as
\begin{equation}
\label{relh1}h_3(u,v,\widetilde{\rho })=\frac 1{\sinh ^2u+\sin ^2v}\frac{%
\left[ 1-\frac{r_g}{4\widetilde{\rho }}\right] ^2}{\left[ 1+\frac{r_g}{4%
\widetilde{\rho }}\right] ^6}.
\end{equation}
In order to define the $h_4$ coefficient solving the Einstein equations, for
simplicity, with a diagonal energy--momentum d--tensor for vanishing pressure,
 we must solve the equation (\ref{heq1}) which transforms into a linear
equation (\ref{heq1a}) if $\Upsilon _1=0.$ In our case $s\left( u,v,\varphi
\right) =\beta ^{-1}\left( u,v,\varphi \right) ,$ where $\beta =\left(
\partial h_4/\partial \varphi \right) /h_4,$ must be a solution of
$$
\frac{\partial s}{\partial \varphi }+\frac{\partial \ln \sqrt{\left|
h_3\right| }}{\partial \varphi }\ s=\frac 12.
$$
After two integrations (see \cite{kamke}) the general solution for $%
h_4(u,v,\varphi ),$ is
\begin{equation}
\label{relh1a}h_4(u,v,\varphi )=a_4\left( u,v\right) \exp \left[
-\int\limits_0^\varphi F(u,v,z)\ dz\right] ,
\end{equation}
where%
\begin{eqnarray}
F(u,v,z)&=& (\sqrt{|h_3(u,v,z)|}[s_{1(0)}\left( u,v\right) \nonumber \\
&{}& +
\frac 12\int\limits_{z_0\left( u,v\right) }^z\sqrt{|h_3(u,v,z)|}dz])^{-1},
 \nonumber
\end{eqnarray}
$s_{1(0)}\left( u,v\right) $ and $z_0\left( u,v\right) $ are some functions
of necessary smooth class. We note that if we put $h_4=a_4(u,v)$ the
equations (\ref{einsteq3b}) are satisfied for every $h_3=h_3(u,v,\varphi ).$

Every d--metric (\ref{rel1}) with coefficients of type (\ref{relh1h}), (\ref
{relh1}) and (\ref{relh1a}) solves the Einstein equations (\ref{einsteq3a}%
)--(\ref{einsteq3d}) with the diagonal momentum d--tensor
$$
\Upsilon _\beta ^\alpha =diag\left[ 0,0,-\varepsilon =-m_0,0\right] ,
$$
when $r_g=2\kappa m_0;$ we set the light constant $c=1.$ If we choose
$$
a_4\left( u,v\right) =\frac{\sinh ^2u\ \sin ^2v}{\sinh ^2u+\sin ^2v}
$$
our solution is conformally equivalent (if not considering the time--time
component) to the hypersurface metric (\ref{hsuf1}). The condition of
vanishing of the coefficient (\ref{relh1}) parametrizes the rotation
ellipsoid for the horizon%
$$
\frac{\widetilde{x}^2+\widetilde{y}^2}{\sigma ^2-1}+\frac{\widetilde{z}^2}{%
\sigma ^2}=\left( \frac{r_g}4\right) ^2,
$$
where the radial coordinate is redefined via relation\ $\widetilde{r}=%
\widetilde{\rho }\left( 1+\frac{r_g}{4\widetilde{\rho }}\right) ^2. $ After
multiplication on the conformal factor
$$
\left( \sinh ^2u+\sin ^2v\right) \left[ 1+\frac{r_g}{4\widetilde{\rho }}%
\right] ^4,
$$
approximating $g_1(u,v)=\sin ^2v\approx 0,$ in the limit of locally
isotropic spherical symmetry,%
$$
\widetilde{x}^2+\widetilde{y}^2+\widetilde{z}^2=r_g^2,
$$
the d--metric (\ref{rel1}) reduces to
$$
ds^2=\left[ 1+\frac{r_g}{4\widetilde{\rho }}\right] ^4\left( d\widetilde{x}%
^2+d\widetilde{y}^2+d\widetilde{z}^2\right) -\frac{\left[ 1-\frac{r_g}{4%
\widetilde{\rho }}\right] ^2}{\left[ 1+\frac{r_g}{4\widetilde{\rho }}\right]
^2}dt^2
$$
which is just the Schwazschild solution with the redefined radial coordinate
when the space component becomes conformally Euclidean.

So, the d--metric (\ref{rel1}), the coefficients of N--connection being
solutions of (\ref{einsteq3c}) and (\ref{einsteq3d}), describe a static 4D
solution of the Einstein equations when instead of a spherical symmetric
horizon one considers a locally anisotropic deformation to the hypersurface
of rotation elongated ellipsoid.

\vskip0.2cm

\paragraph{Flattened rotation ellipsoid coordinates}

${~}$
\\ ${\qquad}$ In a similar fashion we can construct a static 4D black hole
solution with the horizon parametrized by a flattened rotation ellipsoid
\cite{korn},
$$
\frac{\widetilde{x}^2+\widetilde{y}^2}{1+\sigma ^2}+\frac{\widetilde{z}^2}{%
\sigma ^2}=\widetilde{\rho }^2,
$$
where $\sigma \geq 0$ and $\sigma =\sinh u.$

The space 3D special coordinate system is defined%
\begin{eqnarray}
\widetilde{x} &=&\widetilde{\rho}\cosh u\sin v\cos \varphi ,\
\widetilde{y}=\widetilde{\rho}\cosh u\sin v\sin \varphi ,\ \nonumber \\
\widetilde{z} &=& \widetilde{\rho} \sinh u\cos v, \nonumber
\end{eqnarray}
where $0\leq u<\infty ,\ 0\leq v\leq \pi ,\ 0\leq \varphi <2\pi .$

The hypersurface metric is
\begin{eqnarray}
g_{uu} &=& g_{vv}=\widetilde{\rho}^2\left( \sinh ^2u+\cos ^2v\right) ,
 \nonumber \\
g_{\varphi \varphi } &=&\widetilde{\rho}^2\sinh ^2u\cos ^2v.
 \nonumber
\end{eqnarray}
In the rest the black hole solution is described by the same formulas as in
the previous subsection but with respect to new canonical coordinates for
flattened rotation ellipsoid.

\subsubsection{Cylindrical, Bipolar and Toroidal Configurations}

We consider a d--metric of type (\ref{dmetr4}). As a coefficient for
h--metric we choose $g_1(\chi ^1,\chi ^2)=\left( \cos \chi ^2\right) ^{2}$
which solves the Einstein equations (\ref{einsteq3a}). The energy momentum
d--tensor is chosen to be diagonal, $\Upsilon _\beta ^\alpha
=diag[0,0,-\varepsilon ,0]$ with $\varepsilon \simeq m_0=\int m_{(lin)}dl,$
where $\varepsilon _{(lin)}$ is the linear 'mass' density. The coefficient $%
h_3\left( \chi ^i,z\right) $ will be chosen in a form similar to (\ref{relh1}%
),%
$$
h_3\simeq \left[ 1-\frac{r_g}{4\widetilde{\rho }}\right] ^2/\left[ 1+\frac{%
r_g}{4\widetilde{\rho }}\right] ^6
$$
for a cylindrical elliptic horizon. We parametrize the second v--component
as $h_4=a_4(\chi ^1,\chi ^2)$ when the equations (\ref{einsteq3b}) are
satisfied for every $h_3=h_3(\chi ^1,\chi ^2,z).$

We note that in this work we only proof the existence of the mentioned
 type horizon configurations. The exact solutions and physics of
  so--called   ellipsoidal black holes, black torus, black cylinders and
  black disks with, or not, local anisotropy will be examined in
  \cite{vacarubt}.

\vskip0.2cm

\paragraph{Cylindrical coordinates:}

${~}$
\\ $\qquad$
Let us construct a solution of the Einstein equation with the horizon
having the symmetry of ellipsoidal cylinder given by hypersurface formula
\cite{korn}
$$
\frac{\widetilde{x}^2}{\sigma ^2}+\frac{\widetilde{y}^2}{\sigma ^2-1}=\rho
_{*}^2,\ \widetilde{z}=\widetilde{z},
$$
where $\sigma \geq 1.$ The 3D radial coordinate $\widetilde{r}$ is to be
computed from $\widetilde{\rho }^2=\rho _{*}^2+\widetilde{z}^2.$

The 3D space coordinate system is defined%
$$
\widetilde{x}=\rho _{*}\cosh u\cos v,\ \widetilde{y}=\rho _{*}\sinh u\sin
v\sin ,\ \widetilde{z}=\widetilde{z},
$$
where $\sigma =\cosh u,\ (0\leq u<\infty ,\ 0\leq v\leq \pi ).$

The hypersurface metric is
\begin{equation}
\label{melcy}g_{uu}=g_{vv}=\rho _{*}^2\left( \sinh ^2u+\sin ^2v\right)
,g_{zz}=1.
\end{equation}

A solution of the Einstein equations with singularity on an ellipse is given
by
\begin{eqnarray}
h_3 &=&
\frac 1{\rho _{*}^2\left( \sinh ^2u+\sin ^2v\right) }\times \frac{\left[
1-\frac{r_g}{4\widetilde{\rho }}\right] ^2} %
{\left[ 1+\frac{r_g}{4\widetilde{\rho }}\right] ^6},  \nonumber \\
h_4 &=& a_4=\frac 1{\rho _{*}^2\left( \sinh ^2u+\sin ^2v\right) },
\nonumber
\end{eqnarray}
where $\widetilde{r}=\widetilde{\rho }\left( 1+\frac{r_g}{4\widetilde{\rho }}%
\right) ^2.$ The condition of vanishing of the time--time coefficient $h_3$
parametrizes the hypersurface equation of the horizon%
$$
\frac{\widetilde{x}^2}{\sigma ^2}+\frac{\widetilde{y}^2}{\sigma ^2-1}=\left(
\frac{\rho _{*(g)}}4\right) ^2,\ \widetilde{z}=\widetilde{z},
$$
where $\rho _{*(g)}=2\kappa m_{(lin)}.$

By multiplying the d--metric on the conformal factor
$$
\rho _{*}^2\left( \sinh ^2u+\sin ^2v\right)
\left[ 1+\frac{r_g}{4\widetilde{\rho }}\right] ^4,
$$
where $r_g=\int \rho _{*(g)}dl$ (the integration is taken along the
ellipse), for $\rho _{*}\rightarrow 1,$ in the local isotropic limit, $\sin
v\approx 0, $ the space component transforms into (\ref{melcy}).

\vskip0.2cm

\paragraph{ Bipolar coordinates:}

${~}$\\ ${\qquad}$
 Let us construct 4D solutions of the Einstein equation with the horizon
having the symmetry of the bipolar hypersurface given by the formula \cite
{korn}%
$$
\left( \sqrt{\widetilde{x}^2+\widetilde{y}^2}-\frac{\widetilde{\rho }}{\tan
\sigma }\ \right) ^2+\widetilde{z}^2=\frac{\widetilde{\rho }^2}{\sin
^2\sigma },
$$
which describes a hypersurface obtained under the rotation of the circles
$$
\left( \widetilde{y}-\frac{\widetilde{\rho }}{\tan \sigma }\right) ^2+%
\widetilde{z}^2=\frac{\widetilde{\rho }^2}{\sin ^2\sigma }
$$
around the axes $Oz$; because $|c\tan \sigma |<|\sin \sigma |^{-1},$ the
circles intersect the axes $Oz.$ The 3D space coordinate system is defined%
\begin{eqnarray}
\widetilde{x} &=&
\frac{\widetilde{\rho}\sin \sigma \cos \varphi }{\cosh \tau -\cos\sigma },
 \qquad
\widetilde{y} =
\frac{\widetilde{\rho}\sin \sigma \sin \varphi }{\cosh\tau -\cos \sigma },
 \nonumber \\
\widetilde{z} & = &\frac{\widetilde{r}\sinh \tau }{\cosh \tau
-\cos \sigma }, \nonumber  \\ &{}&
\left( -\infty <\tau <\infty ,
0\leq \sigma <\pi ,0\leq \varphi <2\pi \right).
 \nonumber
\end{eqnarray}
The hypersurface metric is
\begin{equation}
\label{mbipcy}g_{\tau \tau }=g_{\sigma \sigma }=\frac{\widetilde{\rho }^2}{%
\left( \cosh \tau -\cos \sigma \right) ^2},g_{\varphi \varphi }=\frac{%
\widetilde{\rho }^2\sin ^2\sigma }
{\left( \cosh \tau -\cos \sigma \right) ^2}.
\end{equation}

A solution of the Einstein equations with singularity on a circle is given
by
$$
h_3=\left[ 1-\frac{r_g}{4\widetilde{\rho }}\right] ^2/\left[ 1+\frac{r_g}{4%
\widetilde{\rho }}\right] ^6\mbox{ and }h_4=a_4=\sin ^2\sigma ,
$$
where $\widetilde{r}=\widetilde{\rho }\left( 1+\frac{r_g}{4\widetilde{\rho }}%
\right) ^2.$ The condition of vanishing of the time--time coefficient $h_3$
parametrizes the hypersurface equation of the horizon%
$$
\left( \sqrt{\widetilde{x}^2+\widetilde{y}^2}-\frac{r_g}2\ c\tan \sigma
\right) ^2+\widetilde{z}^2=\frac{r_g^2}{4\sin ^2\sigma },
$$
where $r_g=\int \rho _{*(g)}dl$ (the integration is taken along the circle),
$\rho _{*(g)}=2\kappa m_{(lin)}.$

By multiplying the d--metric on the conformal factor
\begin{equation}
\label{confbip}\frac 1{\left( \cosh \tau -\cos \sigma \right) ^2}\left[ 1+%
\frac{r_g}{4\widetilde{\rho }}\right] ^4,
\end{equation}
for $\rho _{*}\rightarrow 1,$ in the local isotropic limit, $\sin v\approx
0, $ the space component transforms into (\ref{mbipcy}).

\vskip0.2cm

\paragraph{ Toroidal coordinates:}

${~}$\\ ${\qquad}$
Let us consider solutions of the Einstein equations with toroidal
symmetry of horizons. The hypersurface formula of a torus is \cite{korn}%
$$
\left( \sqrt{\widetilde{x}^2+\widetilde{y}^2}-\widetilde{\rho }\ c\tanh
\sigma \right) ^2+\widetilde{z}^2=
\frac{\widetilde{\rho }^2}{\sinh ^2\sigma }.
$$
The 3D space coordinate system is defined%
\begin{eqnarray}
\widetilde{x} &=&
\frac{\widetilde{\rho}\sinh \tau \cos \varphi }{\cosh \tau -\cos\sigma },
 \qquad
\widetilde{y} = \frac{\widetilde{\rho}\sin \sigma \sin \varphi }{\cosh
\tau -\cos \sigma }, \nonumber \\
\widetilde{z} &=& \frac{\widetilde{\rho}\sinh \sigma }{\cosh
\tau -\cos \sigma }, \nonumber \\ &{}&
\left( -\pi <\sigma <\pi , 0\leq \tau <\infty ,0\leq \varphi <2\pi \right) .
 \nonumber
\end{eqnarray}
The hypersurface metric is
\begin{equation}
\label{mtor}g_{\sigma \sigma }=g_{\tau \tau }=\frac{\widetilde{\rho }^2}{%
\left( \cosh \tau -\cos \sigma \right) ^2},
g_{\varphi \varphi }=\frac{\widetilde{\rho }^2\sin ^2\sigma }
{\left( \cosh \tau -\cos \sigma \right) ^2}.
\end{equation}

This, another type of solution of the Einstein equations with singularity on
a circle, is given by
$$
h_3=\left[ 1-\frac{r_g}{4\widetilde{\rho }}\right] ^2/\left[ 1+\frac{r_g}{4%
\widetilde{\rho }}\right] ^6\mbox{ and }h_4=a_4=\sinh ^2\sigma ,
$$
where $\widetilde{r}=\widetilde{\rho }\left( 1+\frac{r_g}{4\widetilde{\rho }}%
\right) ^2.$ The condition of vanishing of the time--time coefficient $h_3$
parametrizes the hypersurface equation of the horizon%
$$
\left( \sqrt{\widetilde{x}^2+\widetilde{y}^2}-\frac{r_g}{2\tanh \sigma }%
c\right) ^2+\widetilde{z}^2=\frac{r_g^2}{4\sinh ^2\sigma },
$$
where $r_g=\int \rho _{*(g)}dl$ (the integration is taken along the circle),
$\rho _{*(g)}=2\kappa m_{(lin)}.$

By multiplying the d--metric on the conformal factor (\ref{confbip}), for $%
\rho _{*}\rightarrow 1,$ in the local isotropic limit, $\sin v\approx 0, $
the space component transforms into (\ref{mtor}).

\subsection{Disks with Local Anisotropy}

The d--metric is of type (\ref{rel1})
\begin{eqnarray}
\label{diskla}\delta s^2 &= & g_1(\rho,\zeta)d\rho ^2+d\zeta ^2 + \\
 &{} & h_3\left( \rho,\zeta ,\varphi \right)
\left( \delta t\right) ^2+h_4\left( \rho,\zeta,\varphi \right) \left( \delta
\varphi ' \right) ^2, \nonumber
\end{eqnarray}
where the 4D coordinates are parametrized $x^1=\rho $ (the coordinate
radius), $x^2=\zeta ,y^3=t,y^4=\widetilde{\varphi }$ (like for the disk
solution in general relativity \cite{neugm}) and $\delta t$ and $\delta
\widetilde{\varphi }$ are N--elongated differentials). One uses also primed
coordinates given with respect to corotating frame of reference, $\rho
^{\prime }=\rho ,\zeta ^{\prime }=\zeta ,\varphi ^{\prime }=\varphi -\Omega
t,t^{\prime }=t,$ where $\Omega $ is the angular velocity as measured by an
observer at $\infty .$ The h--coordinates run respectively values $0\leq
\rho <\infty $ and $-\infty <\zeta <\infty .$ We consider a disk defined by
conditions $\zeta =0$ and $\rho \leq \rho _0.$

As a particular solution (\ref{btzlh4}) for the h--metric we choose the
coefficient
\begin{equation}
\label{disk1}g_1(\rho ,\zeta )=\left( \cos \zeta \right) ^{2}.
\end{equation}
The explicit form of coefficients $h_3$ and $h_4$ are defined by using
functions
\begin{equation}
\label{coefdisk}A=\rho ^2\exp [2\left( U-k\right) ],\ B=-\exp \left(
4U\right)
\end{equation}
and
$$
\widetilde{\varphi }=\varphi -\frac{Ba}{A-Ba^2}t
$$
where $U,k,$ and $a$ are some functions on $\left( \rho ,\zeta ,\varphi
\right) .$ For the locally isotropic disk solutions
 we  consider only dependencies on
 $\left( \rho ,\zeta\right) ,$ in this case we shall write
$$
U=U_0\left( \rho ,\zeta \right) ,k=k_0\left( \rho ,\zeta \right)
,a=a_0\left( \rho ,\zeta \right)
$$
and
$$
A=A_0\left( \rho ,\zeta \right) ,B=B_0\left( \rho ,\zeta \right),
$$
where the values with the index $0$ are computed by using the
 Neugebauer and Meinel disk solution \cite{neugm}.
The (energy) mass density is taken%
$$
\varepsilon \left( \rho ,\zeta ,\varphi \right) =\delta \left( \zeta \right)
\exp \left( U-k\right) \sigma _p\left( \rho ,\varphi \right) ,
$$
where $\sigma _p\left( \rho ,\varphi \right) $ is the (proper) surface mass
density which is (in the la--case) non--uniformly distributed on the disk;
for locally isotropic distributions $\sigma _p=\sigma _p\left( \rho \right).$
 After the problem is solved one computes $\sigma _p$ as%
$$
\sigma _p=\frac 1{2\pi }e^{U-k}\frac{\partial U^{\prime }}{\partial \zeta }%
\mid _{\zeta =0^{+}}.
$$

The time--time component $h_3$ is chosen in the form
$$
h_3\left( \rho ,\zeta ,\varphi \right) =-\frac{AB}{A-Ba^2}
$$
and, for simplicity, we state the second v--component of d--metric $h_4$ to
depend only h--coordinates as
$$
h_4=a_4\left( \rho ,\zeta \right) =A_0-B_0\ a_0^2.
$$

As in the locally isotropic case one introduces the complex
Ernst potential%
$$
f\left( \rho ,\zeta ,\varphi \right) =e^{2U\left( \rho ,\zeta ,\varphi
\right) }+ib\left( \rho ,\zeta ,\varphi \right) ,
$$
which depends additionally on coordinate $\varphi .$ If the real and
imaginary part of this potential are defined the coefficients (\ref
{coefdisk}) are computed
\begin{eqnarray}
a\left( \rho ,\zeta ,\varphi \right) & = &\int\limits_0^\rho \tilde \rho
e^{-4U}b,_\zeta d\tilde \rho , \nonumber \\
k\left( \rho ,\zeta ,\varphi \right)
 & = & \int\limits_0^\rho \tilde \rho [U,_{\tilde \rho }^2-U,_
 \zeta ^2+\frac 14e^{-4U}(b,_{\tilde \rho }^2-b,_\zeta
^2)]d\tilde \rho .  \nonumber
\end{eqnarray}
[In the integrands, one has $U=U(\tilde \rho ,\zeta ,\varphi )$ and $b=b(%
\tilde \rho ,\zeta ,\varphi )$.]

The Ernst potential is computed as in the locally isotropic limit with that
difference that the values also depend on angular parameter $\varphi ,$%
\begin{equation}
\label{ernst}f=\exp \left\{ \int\limits_{K_1}^{K_a}\frac{K^2dK}Z%
+\int\limits_{K_2}^{K_b}\frac{K^2dK}Z-v_2\right\} ,
\end{equation}
with
\begin{eqnarray}
Z &= &\sqrt{(K+iz)(K-i\bar z)(K^2-K_1^2)(K^2-K_2^2)}, \nonumber \\
K_1 &= &
\rho _0\sqrt{\frac{i-\mu }\mu }\quad (\Re K_1<0),\quad K_2=-\bar K_1,
 \nonumber
 \end{eqnarray}
where $\Re$ denotes the real part. The real (positive) parameter $\mu $ is
given by
$$
\mu =2\Omega ^2\rho _0^2e^{-2V_0}
$$
where $V_0=const.$ The upper integration limits $K_a$ and $K_b$ in (\ref
{ernst}) are calculated from
\begin{equation}
\label{jacobi}\int\limits_{K_1}^{K_a}\frac{dK}Z+\int\limits_{K_2}^{K_b}\frac{%
dK}Z=v_0,\quad \int\limits_{K_1}^{K_a}\frac{KdK}Z+\int\limits_{K_2}^{K_b}%
\frac{KdK}Z=v_1,
\end{equation}
where the functions $v_0$, $v_1$ and $v_2$ in (\ref{jacobi}) and (\ref{ernst}%
) are given by
\begin{eqnarray}
\label{v}v_0 &=& \int\limits_{-i\rho _0}^{+i\rho _0}\frac H{Z_1}dK,\quad
v_1=\int\limits_{-i\rho _0}^{+i\rho _0}\frac H{Z_1}KdK, \\
v_2 &= &\int\limits_{-i\rho _0}^{+i\rho _0}\frac H{Z_1}K^2dK, \nonumber
\end{eqnarray}
\begin{eqnarray}
H &=&\frac{\mu \ln \left[ \sqrt{1+\mu ^2(1+K^2/\rho _0^2)^2}+
\mu (1+K^2/\rho _0^2)\right] }{\pi i\rho _0^2
\sqrt{1+\mu ^2(1+K^2/\rho _0^2)^2}} \nonumber \\
 (\Re H &=& 0), \nonumber
\end{eqnarray}
$$
Z_1=\sqrt{(K+iz)(K-i\bar z)},
$$
where $\Re Z_1<0$ for $\rho$ and $\zeta$ outside the disk. In (\ref{v}) one
has to integrate along the imaginary axis. The integrations from $K_1$ to $%
K_a$ and $K_2$ to $K_b$ in (\ref{ernst}) and (\ref{jacobi}) have to be
performed along the same paths in the two--seethed Riemann surface
associated with $Z(K)$. The problem of finding $K_a$ and $K_b$ from (\ref
{jacobi}) is a special case of Jacobi's inversion problem.

So, we have constructed a locally anisotropic generalization of the
Neugebauer--Meinel \cite{neugm}
disk solution in general relativity, with an additional dependence
 on angle $\varphi$. In the locally
isotropic limit, $g_1=\left( \cos \zeta \right) ^{2}\approx 1,$ when the
d--metric (\ref{diskla}) is conformally equivalent (with the factor $\exp
[2(U_0\left( \rho ,\zeta \right) -k_0\left( \rho ,\zeta \right) )]$) to the
disk solution from \cite{neugm}).

\subsection{Locally Anisotropic generalizations of the Schwarzschild and
Kerr solutions}

\subsubsection{A Schwarzschild like la--solution}

The d--metric of type (\ref{rel1}) is taken
\begin{eqnarray}
\label{schla}\delta s^2 &= &g_1(\chi ^1,\theta )d(\chi ^1)^2+d\theta^2+ \\
 &{}&
h_3\left( \chi ^1,\theta ,\varphi \right) \left( \delta t\right)^2
+h_4\left( \chi ^1,\theta ,\varphi \right)
 \left( \delta \varphi \right)^2, \nonumber
\end{eqnarray}
where on the horizontal subspace $\chi ^1=\rho /r_a$ is the undimensional 
radial coordinate (the constant $r_a$ will be defined below), $\chi
^2=\theta $ and in the vertical subspace $y^3=z=t$ and $y^4=\varphi .$ The
energy--momentum d--tensor is taken to be diagonal $\Upsilon _\beta ^\alpha
=diag[0,0,-\varepsilon ,0].$ The coefficient $g_1$ is chosen to be a
solution of type (\ref{btzlh4})%
$$
g_1\left( \chi ^1,\theta \right) =\cos ^2\theta .
$$
For
$h_4=\sin ^2\theta$ and 
$h_3(\rho ) =-[1-r_a/4\rho ]^2 / [1+r_a/4\rho ]^6,$ 
where  $r= $ $\rho (1+\frac{r_g}{4\rho })^2,$ $ r^2=x^2+y^2+z^2,$ $%
r_a\dot =r_g$ is the Schwarzschild gravitational radius, the d--metric (\ref
{schla}) describes a la--solution of the Einstein equations which is
conformally equivalent, with the factor $\rho ^2\left( 1+r_g/4\rho \right)
^2,$ to the Schwarzschild solution (written in coordinates $\left( \rho
,\theta ,\varphi ,t\right) ),$ embedded into a la--background given by
non--trivial values of $q_i(\rho ,\theta ,t)$ and $n_i(\rho ,\theta ,t).$ In
the anisotropic case we can extend the solution for anisotropic (on angle $%
\theta )$ gravitational polarizations of point particles masses, $m=m\left(
\theta \right) ,$ for instance in elliptic form, when
$$
r_a\left( \theta \right) =\frac{r_g}{\left( 1+e\cos \theta \right) }
$$
induces an ellipsoidal dependence on $\theta $ of  the radial coordinate,%
$$
\rho =\frac{r_g}{4\left( 1+e\cos \theta \right) }.
$$
We can also consider arbitrary solutions with $r_a=r_a\left( \theta ,t\right)
$ of oscillation type, $r_a\simeq \sin \left( \omega _1t\right) ,$ or
modelling the mass evaporation, $r_a\simeq \exp [-st],s=const>0.$

So, fixing a physical solution for $h_3(\rho ,\theta ,t),$ for instance,
$$
h_3(\rho ,\theta ,t)=-\frac{\left[ 1-r_a\exp [-st]/4\rho \left( 1+e\cos
\theta \right) \right] ^2}{\left[ 1+r_a\exp [-st]/4\rho \left( 1+e\cos
\theta \right) \right] ^6},
$$
where $e=const<1,$ and computing the values of $q_i(\rho ,\theta ,t)$ and $%
n_i(\rho ,\theta ,t)$ from (\ref{einsteq3c}) and (\ref{einsteq3d}),
corresponding to given $h_3$ and $h_4,$ we obtain a la--generalization of
the Schwarzschild metric.

We note that fixing this type of anisotropy,  in the locally isotropic limit
we obtain not just the Schwarzschild metric but a conformally transformed
one, multiplied on the factor $1/\rho ^2\left( 1+r_g/4\rho \right) ^4.$

\subsubsection{A Kerr like la--solution}

The d--metric is of type (\ref{diskla}) is taken
\begin{eqnarray}
\delta s^2 &=& g_1\left( r/r_g,\theta \right) dr^2+d\theta ^2+ \nonumber \\
 &{}&
h_3\left(r/r_g,\theta ,\widetilde{\varphi }\right) \left( \delta t\right)^2
+h_4\left( r/r_g,\theta ,\widetilde{\varphi }\right) \left( \delta
\widetilde{\varphi }\right) ^2. \nonumber
\end{eqnarray}
In the locally isotropic limit this metric is conformally equivalent to the
Kerr solution, with the factor $r_{\circ}^2=r^2+a_{\circ}^2,$ $a_{\circ}
=const$ is
associated to the rotation momentum, if
$$
g_1^{[i]}=1/\bigtriangleup \left( r\right) ,
$$
where $\bigtriangleup (r)=r^2-rr_g+a_{\circ}^2,$ $r_g$ is the gravitational
radius and the index $[i]$ points to locally isotropic values,
$$
h_4^{[i]}=A\left( r,\theta \right) ,
$$
where $
A\left( r,\theta \right) =\frac{\sin ^2\theta }{r_{\circ}^2}
\left( r^2+a_{\circ}^2+\frac{rr_ga_{\circ}^2}{r_{\circ}^2}
\sin ^2\theta \right)$
and
$$
h_3^{[i]}=-\left( \frac{Q^2}A+B\right) ,
$$
where $Q=\frac{rr_ga_{\circ}}{r_{\circ}^4}\sin ^2\theta$  and
$B=\frac 1{r_{\circ}^2}\left( 1-\frac{rr_g}{r_{\circ}^2}\right) .$ 
The tilded angular variable $\widetilde{\varphi }$ is introduced with the
aim to get a diagonal d--metric, $
\widetilde{\varphi }=\varphi -\frac QAt.$

A locally anisotropic generalization is to be found if we consider, for
instance, that $r_g\rightarrow r_g\left( \theta \right) $ is defined by an
anisotropic mass $\tilde{m}(\theta)$
 and the locally isotropic values with $%
r_g=cons$ are changed into those with variable $r_g\left( \theta \right) .$
The d--metric coefficient $h_4\left( r,\theta ,\widetilde{\varphi }\right) $
and the corresponding N--connection components are taken as to solve the
equations (\ref{einsteq3b}) and (\ref{einsteq3c}).

\section{Some Additional Examples}
\subsection{Two--soliton locally anisotropic solutions}

Instead of one soliton solutions  we can also consider
la--deformations of multi--soliton configurations (as a review see \cite
{solitons}). In this Subsection we give an example of anholonomic, for
simplicity, two--soliton configuration in general relativity. The d--metric
to be constructed is of Class 1 with the h--component being a solution
  (type (\ref{sgla1})) of the equation (\ref{sgla}) for $\epsilon =-1$ and
 $x^i=(t,x).$

The horizontal component of d--metric is induced, via a conformal transform
 (\ref{conf3}), from the so--called two soliton 2D Lorentz metric
(here we follow the denotations from \cite{gegenberg} being adapted to
 locally anisotropic constructions)
$$
d\widetilde{s}^2=-2\frac{FG}{F^2+G^2}dt^2+\frac{G^2-F^2}{F^2+G^2}dx,
$$
where for the two soliton solution%
\begin{eqnarray}
F &=&\cot \mu \sinh \left[ \widetilde{m}\sin \mu \gamma
 \left( t+vx\right)\right] , \nonumber \\
G &=& \sinh \left[ \widetilde{m}\cos \mu \gamma
 \left( x-vt\right) \right]  \nonumber
\end{eqnarray}
or, for the soliton--anti-soliton solution,%
\begin{eqnarray}
F &=&\cot \mu \cosh \left[ \widetilde{m}\sin \mu \gamma
\left( t+vx\right)\right] , \nonumber \\
G &=& \cosh \left[ \widetilde{m}\cos \mu \gamma
 \left( x-vt\right) \right] , \nonumber
 \end{eqnarray}
with constant parameters $\mu ,\gamma $ and $v.$  The function
$$
u\left( t,x\right) =4\tan ^{-1}\left( F/G\right)
$$
is a solution of 2D Euclidean sine--Gordon equation%
$$
\partial _t^2u+\partial _x^2u=\widetilde{m}^2\sin u.
$$

The locally anisotropic deformation is described by a la--dilaton fild $%
\omega \left( x^i\right) $ chosen to solve the Poisson equation (\ref
{poisson}) with the source $\rho $ (\ref{rho}) computed by using the two
soliton function $u\left( t,x\right) .$ In consequence, the h--metric is of
the form
\begin{equation}
\label{hmmsol}g=\exp [\omega \left( x^i\right) ]\times \left[ -2\frac{FG}{%
F^2+G^2}dt^2+\frac{G^2-F^2}{F^2+G^2}\right] dx.
\end{equation}

The next step is the construction of a soliton like v--metric. Let, for
simplicity,
\begin{equation}
\label{h3mmsol}h_3=a_3\left( x^i\right) =\exp [\omega \left( x^i\right) ]%
\frac{G^2-F^2}{F^2+G^2}
\end{equation}
and $h_4=h_4\left( x^i,z\right) $ is to be defined by the equation (\ref{heq}%
), which for $\partial h_3/\partial z=0$ transforms into
\begin{equation}
\label{h4mmeq}h_4\frac{\partial ^2h_4}{\partial z^2}-\frac 12\left( \frac{%
\partial h_4}{\partial z}\right) ^2-\frac{k\Upsilon _1}2a_3\left( x^i\right)
\left( h_4\right) ^2=0,
\end{equation}
where  we consider a diagonal energy--momentum d--tensor $%
\Upsilon _\beta ^\alpha =diag[-\varepsilon ,0,0,0].$ Introducing a new
variable $h_4=\xi ^2,$ the equation (\ref{h4mmeq}) transform into a linear
second order differential equation on $z$ when coordinates
 $x^i$ are treated as parameters,%
$$
\frac{\partial ^2\xi }{\partial z^2}+\lambda \left( x^i\right) \xi =0,
$$
where $\lambda \left( x^i\right) =\kappa \varepsilon a_3\left( x^i\right)
/2. $ The general solution\\ $\xi (x^i,z)$ is
$$
\xi = \left\{
\begin{array}{rcl}
c_1 \cosh ( z\sqrt{|\lambda |}) + c_2 \sinh ( z\sqrt{|\lambda |}) & , &
\lambda <0; \\
c_1 + c_2 z & , & \lambda =0; \\
c_1 \cos ( z\sqrt{\lambda }) +c_2 \sin ( z\sqrt{\lambda }) & , & \lambda >0,
\end{array}
\right.
$$
where $c_1$ and $c_2$ are some functions on $x^i.$

The coefficients $h_3=a_3(x^i)$ (\ref{h3mmsol}) and $h_4=\xi ^2\left(
x^i,z\right) $ define a h--metric induced by a 2D two soliton equation. The
complete d--metric solving the Einstein equations (\ref{einsteq2}) is
defined by considering v--coefficients of type (\ref{hmmsol}).

\subsection{Kadomtsev--Petviashvily structures and
non--diagonal energy--momentum d--tensors}

Such structures, for diagonal energy--momentum d--tensors and vacuum
Einstein equations, where proven to exist in Subsection IVB, paragraph 2.
 Here we show
 that another type of three dimensional soliton structures could be
 generated by
  nondiagonal components $\Upsilon _{31}$ and $\Upsilon _{32}.$

For $\Upsilon _1^1=\Upsilon _2^2=0$ every function $h_4=a_4\left( x^i\right)
$ solves the v--component of Einstein equations (\ref{einsteq3b}). Let us
consider a function $h_3=h_3\left( x^i,z\right) .$ If the anholonomic
constraints on the system of reference are imposed by
N--connection coefficients $N_i^3=q_i,$ when
 $$
q_i = -2\kappa \Upsilon _{3i}h_3\left[ h_3\left( h_3^{*}\right) ^2+\epsilon
\left( \dot h_3+6h_3h_3^{\prime }+h_3^{\prime \prime \prime }\right)
^{\prime }\right] ^{-1}, 
 $$
where $\epsilon =\pm 1,$ the system of equations (\ref{einsteq3c}) reduces
to the Kadomtsev--Petviashvili equation for $h_3,$%
$$
h_3^{**}+\epsilon \left( \dot h_3+6h_3h_3^{\prime }+h_3^{\prime \prime
\prime }\right) ^{\prime }=0.
$$
The solution of Einstein equations is to completed by considering some
functions $N_i^4=n_i$ satisfying (\ref{einsteq3d}) and a h--metric $%
g_{ij}(x^k)$ solving (\ref{einsteq3a}).

\subsection{Anholonomic soliton like vacuum configurations}

The main result of Belinski--Zakharov--Maison works \cite{belinski,mais} was
the proof that vacuum gravitational soliton like structures
 could be defined in the framework of general relativity with
$h_{ab}\left( x^i\right) $ (from 4D metric (\ref{belmais})) being a solution
of a generalized type of sine--Gordon equations. The
 function $f(x^i)$ (from
(\ref{belmais})) is to be determined by some integral relations
 after the components $h_{ab}\left( x^i\right) $ have been constructed.

By reformulating the problem of definition of soliton like integral
varieties of vacuum Einstein equations from the viewpoint of anholonomic
frame structures, there are possible further generalizations and
 constructions of new classes of solutions.

For vanishing energy--momentum d--tensors the Einstein equations (\ref
{einsteq3a})--(\ref{einsteq3d}) transform into
\begin{eqnarray}
 &2(g_1^{^{\prime \prime }}&+{\ddot g}_2)
 -\frac 1{g_2}\left( {\dot g}_2^2 +
 g_1^{\prime }g_2^{\prime }\right)-\frac 1{g_1}\left( g_1^{\prime \ 2}+%
\dot g_1\dot g_2\right) = 0; \label{einsteq4a} \\
 &{}& \nonumber  \\
&h_4^{**}& -\frac 1{2h_4}(h_4^{*})^2-\frac 1{2h_3}h_3^{*}h_4^{*} = 0;
 \label{einsteq4b}
\end{eqnarray}
\begin{eqnarray}
2q_1h_4\left[ \left( \frac{h_3^{*}}{h_3}\right) ^2-\frac{h_3^{**}}{h_3}+%
\frac{h_4^{*}}{2h_4^{\ 2}}-\frac{h_3^{*}h_4^{*}}{2h_3h_4}\right] + &{} &
\nonumber \\
\left[ \frac{\dot h_4}{h_4}\ h_4^{*}-2\dot h_4^{*}+\frac{\dot h_3}{h_3}\
h_4^{*}\right] &=& 0, \label{einsteq4c} \\
2q_2h_4\left[ \left( \frac{h_3^{*}}{h_3}\right) ^2-\frac{h_3^{**}}{h_3}+%
\frac{h_4^{*}}{2h_4^{\ 2}}-\frac{h_3^{*}h_4^{*}}{2h_3h_4}\right] +&{} &
\nonumber \\
\left[ \frac{h_4^{\prime }}{h_4}\ h_4^{*}-2h_4^{\prime \ *}+ %
\frac{h_3^{\prime }}{h_3}\ h_4^{*}\right] & = &0; \nonumber
\end{eqnarray}

\begin{equation}
\label{einsteq4d}n_1^{**}=0\mbox{ and }n_2^{**}=0,
\end{equation}
where we suppose that $g_1,g_2,h_3$ and $h_4$ are not zero.

The equation (\ref{einsteq4a}), relates two components and their first and
second order partial derivatives of a diagonal h--metric $g_1(x^i)$ and $%
g_2(x^i).$ We can prescribe one of the components in order to find the
second one by solving a second order partial differential equation. For
instance, we can consider the h--metric to be induced by a soliton--dilaton
solution (like in the Section IV,
 but for vacuum solitons the constants will be
not defined by any components of the energy--momentum d--tensor).

 Let us fix a soliton 2D solution with diagonal auxiliary metric
\begin{eqnarray}
&{}& \widetilde{g}_{ij}  =
diag\{\widetilde{g}_1=\epsilon \sin ^2\left[ v\left(
x^i\right) /2\right] , \widetilde{g}_2  =
 \cos ^2\left[ v\left( x^i\right)/2\right] \},
 \nonumber \\
&{}& \epsilon = \pm 1, \nonumber
\end{eqnarray}
and model the local anisotropy by a la--dilaton field $\omega \left(
x^i\right) $ relating the metric $\widetilde{g}_{ij}$ with the h--components
$g_{ij}$ via a conformal transform (\ref{conf3}). The la--dilaton is to be
found as a solution of the equations (\ref{poisson}) where the source $\rho
\left( x^i\right) $ is computed by using the formula (\ref{rho}).So, we
conclude that vacuum h--metrics can be described by corresponding
soliton--dilaton systems.

The equation (\ref{einsteq4b}) relates two components and their first and
second order partial derivatives on $z$ of a diagonal v--metric $h_3(x^i,z)$
and $h_4(x^i,z)$ which depends on three variables. We also can prescribe one
of these components (for instance, as was shown in details in Section IV B to
be a solution of the Kadomtsev--Patviashvili, or (2+1) dimensional
sine--Gordon equation; the Belinski--Zakharov--Maison solutions can be
considered as some particular case soliton vacuum configurations which
 do not depend on variable $z) $ the second
v--component being defined after solution of the resulted
partial differential equation on $z,$ with the h--coordinates $x^i$ treated
as parameters.

If the values $h_3(x^i,z)$ and $h_4(x^i,z)$ are defined,
 we have algebraic equations
(\ref{einsteq4c}) for calculation of coefficients $q_1(x^i,z)$ and $%
q_2(x^i,z).$ The equations (\ref{einsteq4d}) are satisfied by arbitrary
 $n_1(x^i,z)$ and $n_2(x^i,z)$ depending linearly on the third variable $z.$

\section{Conclusions}

In this paper, we have elaborated a new method of construction of
 exact solutions of the Einstein equations by using   anholonomic
 frames with associated nonlinear connection structures.

We analyzed 4D metrics
$$
ds^2=g_{\alpha \beta }\ du^\alpha du^\beta
$$
when  $g_{\alpha \beta}$  are parametrized by  matrices of type
\begin{equation}
\left[
\begin{array}{cccc}
g_1+q_1{}^2h_3+n_1{}^2h_4 & 0 & q_1h_3 & n_1h_4 \\
0 & g_2+q_2{}^2h_3+n_2{}^2h_4 & q_2h_3 & n_2h_4 \\
q_1h_3 & q_2h_3 & h_3 & 0 \\
n_1h_4 & n_2h_4 & 0 & h_4
\end{array}
\right]   \label{ansatz2}
\end{equation}
with coefficients being some functions of necessary smo\-oth class $%
g_i=g_i(x^j),q_i=q_i(x^j,z),n_i=n_i(x^j,z),$ $h_a=h_a(x^j,z).$  Latin
indices run respectively $i,j,k,...$
$=1,2$ and $a,b,c,...=3,4$ and the local
coordinates are denoted $u^\alpha =(x^i,y^3=z,y^4).$ A metric
 (\ref{ansatz2})  can be diagonalized,
$$
\delta s^2=g_i(x^j)\left( dx^i\right) ^2+
h_a(x^j,z)\left( \delta y^a\right)^2,
$$
with respect to anholonomic frames (\ref{dder}) and (\ref{ddif}), here we
write down only the 'elongated' differentials
$$
\delta z=dz+q_i(x^j,z)dx^i,\
\delta y^4=dy^4+n_i(x^j,z)dx^i.
$$

 The key result of this paper is the proof that for
 the introduced ansatz the Einstein equations simplify substantially
 for 3D and 4D spacetimes, the variables being separated:

\begin{itemize}
 \item
 The equation (\ref{einsteq3a})  with the non--trivial component of the
  Ricci tensor (\ref{ricci1})
  relates two (so--called, horizontal) components of metric
  $g_i$ with the (so--called, vertical) values of the diagonal
  energy--momentum   tensor.  We proved that such components of metric
  could be described
 by soliton--dilaton and black hole like solutions with parameters
 being determined by vertical sources.

 \item
 Similarly,  the equation (\ref{einsteq3b}) with the non--trivial
 component of the  Ricci tensor (\ref{ricci1}) relates two vertical
   components of metric $h_a$ with the horizontal values of the
  diagonal energy--momentum tensor. The vertical coefficients of
  metric could depend on three variables $(x^i,z)$ and this equation
  contains their first and second derivatives on $z,$ the dependence on
  horizontal coordinates $x^i$ being parametric.

 \item
 As to the rest of equations (\ref{einsteq3c}) and (\ref{einsteq3d})
 with corresponding non--trivial Ricci tensors (\ref{ricci3}) and
 (\ref{ricci4}), they form an algebraic system for definition
 of the nonlinear connection coefficients $q_i(x^i,z)$ and second
 order differential equation on $z$ for the nonlinear connection
 coefficients $n_i(x^i,z)$ after the functions $h_a(x^i,z)$ have
 been defined and non--diagonal components of energy--momentum tensor
 are  given.
\end{itemize}

 The Einstein equations consist a system of second order nonlinear
  partial differential equations whose particular solutions are
  selected from the general integral variety by imposing some physical
 motivated conditions on the type of singularities,  horizon hypersurfaces,
  perturbative and/or non--perturbative behavior of background
  configurations, limit correspondences with some well known solutions,
  physical laws, symmetries and so on.

 We investigated the conditions when from the class of solutions
 of 4D and 3D gravitational field equations parametrized by metric
 ansatzs of type (\ref{ansatz2}) we can obtain some locally
 anisotropic generalizations of well known soliton--dilaton, black hole,
 cylinder and disk solutions.

 In this paper we have shown that one can use solutions of generalized
 sine--Gordon equations in two and three dimensions to generate
  4D solutions of Einstein gravity with soliton--dilaton parameters being
 related to 4D energy--momentum values. We have found  a broad class
 of 2D, 3D and 4D black hole configurations with generic local anisotropy.
Our results seem to indicate that there is a deep connection between
 black hole and soliton--dilaton states in gravitational theories
 of lower and 4D dimensions. Via nonlinear superpositions the lower
  dimensional locally anisotropic configurations induce similar structures
  in higher dimensions. We conclude that if
  the former direct applications of the 2D soliton--dilaton--black
 hole models (more naturally treated in the framework of 2D gravity and
 string theory) are  very rough approximations for general relativity,
  after introducing of some well defined principles of  nonlinear
  superposition, the lower dimensional solutions could be  considered as
  some building blocks for construction of non--perturbative
   solutions in four dimensions.

   We presented a series of computations involving the dynamics of
  locally anisotropic gravitational soliton deformations, black hole
   dynamics and constructed exact 4D and 3D solutions of the Einstein
  equations with horizons being (under corresponding dimension)
  of elliptic, rotation ellipsoidal, bipolar, elliptic cylinder and
  toroidal configuration. We showed that such solutions are
  naturally contained in general relativity and defined by
  corresponding anholonomic constraints, anisotropic distributions
   of masses and energy densities and could model some anisotropic
   nonlinear self--gravitational polarizations and renormalizations
   of gravitational and cosmological constants.

  Our approach represents  just a first step in the
  differential geometric and nonlinear analysis of the role that
  solitons and singular configurations with local anisotropy
  plays in 2D, 3D and 4D gravity. The natural developments of our
 approach would be to use nonlinear superpositions to describe
  the semiclassical and quantum dynamics of extremal black
  holes induced from string theory, the corresponding nonequilibrium
    thermodynamics of such black holes.  One would be of interest
  supersymmetric extensions of the method and investigation
  of the mentioned non--perturbative structures in the framework
  of string theory. Work is in progress to address these issues.

\end{document}